\documentclass[12pt]{article}
\usepackage{fullpage,amsmath,amssymb,mathtools,natbib,hyperref}
\usepackage{titling,enumitem}
\usepackage[skip=0pt]{caption} 
\usepackage{multicol}
\usepackage{dirtytalk}
\usepackage{setspace}
\usepackage{datetime,bigints}
\usepackage[dvipsnames]{xcolor}
\usepackage[most]{tcolorbox}
\usdate
\usepackage{multirow,algorithm} 
\usepackage{float}
\usepackage[boxruled,algo2e]{algorithm2e}
\usepackage{arydshln}
\usepackage[section]{placeins}
\usepackage[titletoc,toc,title]{appendix}
\usepackage{fancyvrb} 
\usepackage{listings}
\usepackage{bm}
\usepackage{makecell}
\usepackage{chngcntr}
\usepackage{xcolor,colortbl} 
\newcommand{\notes}[1]{%
    \linespread{0.2}\vspace{0.1em}%
    \captionsetup{justification=justified}%
    \caption*{\footnotesize #1}%
}
\usepackage{hyperref}
\definecolor{lightgray}{gray}{0.9}
\xdefinecolor{gray}{rgb}{0.8,0.8,0.8}
\xdefinecolor{blue}{RGB}{58,95,205}% R's royalblue3; #3A5FCD
\DeclareCaptionFormat{listing}{\rule{\dimexpr\textwidth+17pt\relax}{0.4pt}\vskip1pt#1#2#3}

\usepackage{subcaption}  % For subfigures with captions

\definecolor{gray}{gray}{0.85}
\definecolor{LightCyan}{rgb}{0.88,1,1}

\newcolumntype{a}{>{\columncolor{gray}}c}
\usepackage{xr,caption}
\usepackage{graphicx}
\renewenvironment{thebibliography}[1]{
  \begin{oldthebibliography}{#1}
    \setlength{\itemsep}{0em}
    \setlength{\parskip}{0em}
}
{
  \end{oldthebibliography}
}

\newcommand\nulleq{\stackrel{\mathclap{\normalfont\mbox{$H_0$}}}{=}}

\usepackage{amsthm}
\usepackage{thmtools} 
{
      \theoremstyle{plain}
      \newtheorem{definition}{Definition}
      \newtheorem{theorem}{Theorem}
      \newtheorem{example}{Example}
      \newtheorem{proposition}{Proposition}

      \newtheorem{assumption}{Assumption}
      
  }

\renewcommand{\arraystretch}{1.5}

\def\lQ{\scalebox{-1}[1]{''}}

\makeatletter
\renewenvironment{abstract}{%
    \if@twocolumn
      \section*{\abstractname}%
    \else %% <- here I've removed \small
      \begin{center}%
        {\bfseries \normalsize\abstractname\vspace{\z@}}%  %% <- here I've added \Large
      \end{center} \vspace{-0.5cm}%
      \quotation
    \fi}
    {\if@twocolumn\else\endquotation\fi}
\makeatother

\begin{document}

  %\begin{frontmatter} 
  \title{Unconditional Randomization Tests for Interference}
  \author{Liang Zhong\thanks{Faculty of Business and Economics, The University of Hong Kong, Pokfulam Road, Hong Kong. \newline Email: \href{mailto:samzl@hku.hk}{samzl@hku.hk}, Website:\href{https://samzl1.github.io/}{https://samzl1.github.io/}.  \newline I am deeply grateful to my advisors, Hiroaki Kaido and Jean-Jacques Forneron, for their invaluable guidance and support. I also thank M. Daniele Paserman, Ivan Fernandez-Val, Pierre Perron, Marc Rysman, Xunkang Tian, Stella Hong, Zhongjun Qu, Benjamin Marx, Haoran Pan, Lei Ma, Justin Hong, Qingyuan Chai, Mohsen Bayati, Kirill Ponomarev, Guillaum Pouliot, Jizhou Liu, Julius Owusu, Panos Toulis, Doosoo Kim, Shakeeb Khan, Dmitry Arkhangelsky, Matias Cattaneo, Jos'e R. Zubizarreta, Demian Pouzo, Avi Feller, Xinran Li, Peter Deffebach, Juan Pantano, Toru Kitagawa, and participants at various workshops and seminars for their helpful comments. All errors are my own. }}
   \date{\today}
  \maketitle 

  \begin{abstract}  %\normalsize
Researchers are often interested in the existence and extent of interference between units when conducting causal inference or designing policy. However, testing for interference presents significant econometric challenges, particularly due to complex clustering patterns and dependencies that can invalidate standard methods. This paper introduces the pairwise imputation-based randomization test (PIRT), a general and robust framework for assessing the existence and extent of interference in experimental settings. PIRT employs unconditional randomization testing and pairwise comparisons, enabling straightforward implementation and ensuring finite-sample validity under minimal assumptions about network structure. The method’s practical value is demonstrated through an application to a large-scale policing experiment in Bogot\'a, Colombia \citep{blattman2021}, which evaluates the effects of hotspot policing on crime at the street-segment level. The analysis reveals that increased police patrolling in hotspots significantly displaces violent crime, but not property crime. Simulations calibrated to this context further underscore the power and robustness of PIRT.
  \end{abstract}
  

  \baselineskip=18.0pt
  \thispagestyle{empty}
  \setcounter{page}{0}
  
  %\newpage \tableofcontents
\newpage

\section{Introduction} \label{sec:intro}

When one person receives new information, they often share it with friends. Similarly, when an urban policy is implemented in a neighborhood, its effects can ripple into distant areas. In such situations, the outcome for one unit may depend not only on its own treatment, but also on the treatment assigned to others—a phenomenon known as interference. Understanding the extent of interference is crucial: it refines causal inference and model specification,\footnote{See, for example, \citet{Sacerdote2001}, \citet{Miguel2004}, \citet{Angrist2014}, \citet{paluck2016}, \citet{Jayachandran2017}, \citet{Rajkumar2022}, and \citet{wang2024}.} and it guides efficient resource allocation, especially when interventions are costly \citep{SarahB2018, Eliana2020}.

However, testing for interference presents significant econometric challenges, particularly because complex clustering patterns can render large-sample approximations intractable \citep{morgan2021, blattman2021}. Even in randomized experiments, valid inference may require assumptions beyond simple random assignment \citep{aronow2012, pollmann2023causal}. As a result, recent studies \citep[e.g.,][]{bond2012, blattman2021} have turned to randomization-based approaches, such as the Fisher Randomization Test (FRT), to detect interference. Yet FRTs are generally valid \textit{only} for testing the \textit{sharp null hypothesis} of no treatment effect, which assumes that all potential outcomes remain unchanged under any assignment—an assumption known as imputability \citep{Rosenbaum2007, hudgens2008, athey2018}. In network settings, this sharp null excludes both direct treatment effects and any form of interference. Therefore, when the FRT rejects the sharp null, it may indicate the presence of direct effects, interference, or both, making it impossible to distinguish between these possibilities.

In this paper, I introduce the pairwise imputation-based randomization test (PIRT): an unconditional, design-based framework for detecting and analyzing interference in experimental settings. PIRT treats potential outcomes as fixed and considers random treatment assignment as the sole source of randomness \citep{abadie2020, Abadie}.\footnote{As \citet{blattman2021} note, design-based approaches and randomization inference are particularly well-suited for network contexts with unknown spillover effects.} The core hypothesis tests whether interference exists beyond a specified distance by comparing units’ potential outcomes when they are separated by more than a threshold distance \(\epsilon_s\)
from treated units. The method repeatedly reassigns treatments while holding outcomes fixed, computes targeted test statistics, and derives a \textit{p}-value; a sufficiently small \textit{p}-value provides evidence against the null. PIRT makes minimal assumptions about network structure—making it suitable even for dense or complex networks—and relies solely on random assignment, ensuring validity without further assumptions.

More generally, I define \textit{partially sharp null hypotheses}, in which only a subset of potential outcomes is assumed known across treatment assignments \citep{zhang2023}. Testing for interference is a special case of this framework, as it requires isolating direct treatment effects while assessing whether a unit’s outcome depends on others’ treatment statuses.\footnote{If one is concerned about pre-testing, the proposed method can also be directly used for causal inference, just like a traditional randomization test.} Overall, PIRT is a non-parametric, finite-sample valid, and easily implementable method for testing any partially sharp null hypothesis.\footnote{It is finite-sample exact, meaning the probability of a false rejection in finite samples does not exceed the user-prescribed nominal rate \citep{Guillaum2024}.}

Testing partially sharp null hypotheses with PIRT involves addressing two key technical challenges. First, only a subset of potential outcomes is “imputable”—that is, their values can be inferred from observed data under the partially sharp null hypothesis. For example, under the hypothesis of no peer effects on non-treated units in a social network, outcomes can be imputed only for non-treated units; the outcomes of treated units remain unknown. Second, the set of units with imputable outcomes changes with each treatment assignment, since the identity of non-treated units varies across assignments. Together, these challenges complicate the direct application of traditional randomization inference methods and underscore the need for specialized approaches.

To address the first challenge, I propose a class of test statistics termed \textit{pairwise imputable statistics}, each defined as a function of two treatment assignment vectors. The first assignment determines how units are grouped or compared, while the second identifies those units for which outcomes are imputable. These statistics closely resemble conventional test statistics (as in \citet{Imbens_Rubin_2015}), but are restricted to the subset of imputable units specified by the partially sharp null under both assignments. Despite this restriction, pairwise imputable statistics can accommodate a wide range of commonly used test statistics. For example, a difference-in-means estimator might compare imputable individuals who have treated friends to those who do not: here, the first assignment defines the groupings, paralleling conventional test statistics, while the second determines which individuals are included in the calculation.

To tackle the second challenge, I establish a novel connection between randomization tests for non-sharp null hypotheses and the cross-sectional conformal prediction literature \citep{vovk2018a, barber2021predictive, guan2023conformal}. I construct PIRT \textit{p}-values via pairwise comparisons of two pairwise imputable statistics. In the first, the randomized assignment determines which units are imputable, while the observed assignment defines the groupings and comparisons; in the second, the observed assignment selects the imputable units, with groupings and comparisons determined by the randomized assignment. The validity of this procedure relies on the symmetry of these pairwise comparisons, which is analogous to the conformal lemma of \citet{guan2023conformal}.

To illustrate PIRT’s applicability, I apply the method to a large-scale experiment by \citet{blattman2021}, which evaluated a policing strategy that concentrated resources on high-crime “hotspots” in Bogot\'a, Colombia, using street segments as units. I use PIRT to test for interference—such as crime displacement or deterrence—in nearby neighborhoods.\footnote{This analysis assumes that interactions occur through neighboring units, resulting in potential spillover effects.} While the original authors report significant displacement effects of increased police patrols on property crime but not on violent crime, my PIRT-based analysis yields different conclusions. Specifically, PIRT detects a marginally significant displacement effect on violent crime at the 10\% level, and an insignificant effect on property crime—contrary to the findings of \citet{blattman2021}.\footnote{I also propose a sequential testing procedure that automatically controls the family-wise error rate (FWER) when defining the “neighborhood” of interference.} These results have important implications for welfare analysis and may reshape our understanding of criminal behavior, particularly if more severe violent crime warrants stricter interventions.

A simulation study calibrated to this dataset further demonstrates the strong empirical performance of PIRT compared to existing methods. Specifically, I test for displacement effects, in which interference causes outcomes to “spill over” to neighboring units. At the \( \alpha \) rejection level, PIRT successfully controls type I error rates, maintaining robustness even under worst-case scenarios. In contrast, classical FRT may over-reject under partially sharp null hypotheses. In terms of power, PIRT at the \( \alpha \) level outperforms competing alternatives—an especially important advantage in network analysis, where data collection is costly and interference effects are often subtle \citep{Taylor2018, breza2020}. Nonetheless, there is a trade-off between ease of implementation and conservatism under the null, as PIRT may be conservative in some settings.

\paragraph{Literature Review}
This paper contributes to three main strands of literature. First, it advances the study of network analysis. Following the seminal work of \citet{Manski1993}, a substantial body of research has developed model-based approaches that rely on parametric assumptions \citep[e.g.,][]{Sacerdote2001, bowers2013, Toulis2013, graham2017, dePaula2018}. These approaches typically require the imposition of specific network structures and must contend with the high dimensionality inherent in modeling network interactions. In contrast, my method adopts a nonparametric framework that exploits the null hypothesis to reduce dimensionality, thereby relaxing restrictive parametric assumptions and allowing for greater flexibility in capturing network effects.

Second, this paper contributes to the literature on design-based causal inference under interference. Two principal frameworks have been developed for this setting: the Fisherian and Neymanian perspectives \citep{Li2018}. The Neymanian approach focuses on randomization-based unbiased estimation and variance calculation \citep{hudgens2008, Aronow2017, pollmann2023causal}, typically relying on asymptotic normal approximations and often requiring assumptions such as network sparsity or local interference.\footnote{See also \citet{basse2018}, \citet{viviano2022}, \citet{wang2023designbased}, \citet{VAZQUEZBARE2023}, \citet{Leung2020}, \citet{Leung2022}, and \citet{Bayati2024}.} 

In contrast, this paper adopts the Fisherian perspective, focusing on the detection of causal effects through finite-sample valid, randomization-based tests \citep{dufour2003, lehmann2006, rosenbaum2020design}. Recognizing the limitations of classical Fisher Randomization Tests (FRTs) in the presence of interference, recent literature has developed conditional randomization tests (CRTs), which restrict inference to a subset of units and assignments where the null hypothesis is sharp.\footnote{See, for example, \citet{aronow2012}, \citet{athey2018}, \citet{basse2019b}, \citet{Puelz2022}, \citet{Zhang2021}, \citet{basse2024}, and \citet{hoshino2023}.} However, existing CRTs are often tailored to specific interference structures, such as clustered interference \citep{basse2019b, basse2024}, limiting their generalizability. Moreover, designing effective conditioning events that are both powerful and feasible is challenging, frequently resulting in substantial power loss \citep{Puelz2022}. Implementation of CRTs in settings with general interference can also be computationally intensive. 

This paper builds on these foundations by introducing an alternative approach that applies to a broad class of interference structures, is straightforward to implement, and remains valid even when construction of informative conditioning events is difficult or infeasible. Confidence intervals for causal parameters of interest can subsequently be obtained by inverting these tests. \footnote{Randomization-based methods can also be integrated with model-based frameworks, such as the linear-in-means model \citep{Manski1993}, to increase power or extend applicability beyond randomized experiments while preserving test validity \citep{ding2021, basse2024, Borusyak2023}.}

Finally, this paper contributes to the literature by extending randomization testing to hypotheses that are not fully sharp. While the primary focus is on partially sharp null hypotheses defined via distance measures, the underlying principles of PIRT appear broadly generalizable, potentially extending beyond network contexts. Since \citet{neyman1935} highlighted that FRTs are limited to sharp null hypotheses, subsequent work has developed strategies for testing weak or non-sharp nulls \citep{azeem2025}. For instance, \citet{Ding2016}, \citet{Li2016}, and \citet{ding2020} investigate randomization tests for the null of no average treatment effect, while \citet{caughey2023} considers bounded null hypotheses. \citet{Zhang2021} introduces CRTs for partially sharp nulls, employing approaches similar to those in \citet{athey2018} and \citet{Puelz2022} for time-staggered adoption designs. To the best of my knowledge, PIRT is the first unconditional randomization testing method that accommodates partially sharp null hypotheses, thus broadening the scope of randomization inference in experimental and observational studies.

%\textbf{[Exact tests have existed for some time, one example are Monte Carlo tests, independence testing based on rankings, etc. So I think some of your references look a bit too recent for those who know these things.]}
 
%Most of the classical approaches to causal inference assumed no \textit{interference} \citet{cox1958} between units, which means one unit's treatment does not affect another unit's outcome. However, this assumption is implausible in many settings, such as the scenario related to peer effects and treatment spillovers, so standard approaches often break down when testing interference. Due to the complexity of the large sample approximations to distributions of statistics under the network settings, previous studies looked into \textit{Fisher Randomization Test} (FRT) and proposed testings with exact \textit{p}-values based on the randomization distribution \citet{fisher1925} with a remedy to the jeopardize from the interference -- conditional randomization testing. 

The remainder of the paper is organized as follows. Section~\ref{sec:setup} introduces the general framework, notation, and the null hypothesis of interest. Section~\ref{sec:PIRT} details the PIRT procedure, including the construction of pairwise imputable statistics and the corresponding \textit{p}-value based on pairwise comparisons. Section~\ref{sec:apply} applies the method to a large-scale policing experiment in Bogot\'a, Colombia, while Section~\ref{sec:simu} presents results from a Monte Carlo study calibrated to this setting. Section~\ref{sec:conclude} concludes. Additional empirical and theoretical results, as well as proofs, are provided in the appendix.

\section{Setup and Null Hypothesis of Interest} \label{sec:setup}

Consider $N$ units indexed by $i \in \{1, 2, \dots, N\}$, connected via an undirected network observed by the researcher. My goal is to study the extent of interference between units, as determined by factors such as distance, adjacency, and connection strength. These relationships are summarized by an $N \times N$ proximity matrix $G$. The $(i,j)$-th entry, $G_{i,j} \geq 0$, represents a ``distance measure'' between units $i$ and $j$. This measure can be either continuous or discrete, depending on the application. I normalize $G_{i,i} = 0$ for all $i$, and assume $G_{i,j} > 0$ for all $i \neq j$. The precise definition of ``distance'' is context-specific:\footnote{Researchers may also define distance in product space, particularly for firms selling differentiated products. In this case, units represent products, and $G_{i,j}$ could denote the Euclidean distance in a multi-dimensional space of product characteristics, as in \citet{pollmann2023causal}. Such measures are useful for defining market boundaries, e.g., when merger authorities assess whether two products belong to the same relevant market.}

\begin{example}[Spatial distance] \label{ex:spatial}  
Suppose the units are street segments within a city, as in \citet{blattman2021}. Here, $ G_{i,j} $ represents the spatial distance between street segments $ i $ and $ j $. This setting is typical when local spillovers or interactions are mediated by geographic proximity.
\end{example}

\begin{example}[Network distance] \label{ex:extent}  
Suppose the units are individuals in a social network, such as Facebook users in \citet{bond2012}. In this context, $ G_{i,j} $ measures the network distance between individuals $ i $ and $ j $: for example, $ G_{i,j} = 1 $ if $ i $ and $ j $ are direct friends, $ G_{i,j} = 2 $ if they are friends of friends, and $ G_{i,j} = \infty $ if they are not connected. This formulation accommodates disconnected networks and captures partial interference, such as cluster-level interference \citep{sobel2006, basse2019b}.
\end{example}

\begin{example}[Link intensity] \label{ex:intensity}  
Suppose the units are individuals or entities connected by links of varying strength, such as email correspondents or social contacts. Researchers may observe not only whether two units are linked, but also the intensity of their connection, denoted $ int_{i,j} $ (e.g., frequency of interaction or volume of email correspondence) \citep{Goldenberg2009, bond2012, Rajkumar2022}. Building on the classic study by \citet{mark1973}, one might examine how interference varies across weak and strong ties, defined by this intensity measure. Let $ \bar{int} = \max_{i,j \in \{1, \dots, N\}} int_{i,j} $, and define $ G_{i,j} = \bar{int} - int_{i,j} $. An increase in $ G_{i,j} $ then implies a weaker connection, analogous to Examples~\ref{ex:spatial} and~\ref{ex:extent}.
\end{example}

In this paper, I focus on experimental settings in which treatment assignment is random and follows a known probability distribution $P$, where $P(d) = \Pr(D = d)$ denotes the probability that the treatment assignment vector $D$ equals $d$. Let $X$ represent observed pre-treatment characteristics (e.g., age, gender), which can be used to control for unit heterogeneity. However, I do not attempt to estimate their direct effects on the outcome. The probability distribution $P$ may or may not depend on covariates $X$: in complete or cluster randomization, it does not depend on $X$, while in stratified or matched-pair designs, it does.

I adopt the potential outcomes framework with a binary treatment assignment vector $D = (D_1, \dots, D_N) \sim P$, where $D \in \{0,1\}^N$ and $D_i \in \{0,1\}$ denotes whether unit $i$ is treated. Let $Y(d) = (Y_1(d), \dots, Y_N(d)) \in \mathbb{R}^N$ denote the vector of potential outcomes under treatment assignment $d$, where the potential outcome for unit $i$ is $Y_i(d) = Y_i(d_1, \dots, d_N)$. This notation allows unit $i$'s potential outcome to depend on the treatment assignments of all units, thereby relaxing the classic \textit{Stable Unit Treatment Value Assumption (SUTVA)} of \citet{cox1958} and accommodating settings with spatial or network interference. Throughout, I assume that the proximity matrix $G$ is unaffected by treatment assignment.

The following variables are observed: (1) the realized vector of treatments for all units, denoted by $D^{obs}$; (2) the realized outcomes for all units, denoted by $Y^{obs} \equiv Y(D^{obs}) = (Y_1(D^{obs}), \dots, Y_N(D^{obs}))$; (3) the proximity matrix $G$; (4) the covariates $X$; and (5) the treatment assignment probability distribution $P$. I adopt a design-based inference approach, treating $D$ as random, while $G$, $X$, $P$, and the unknown potential outcome schedule $Y(\cdot)$ are considered fixed throughout. For notational simplicity, these elements will not be treated as arguments of functions in the remainder of the paper.

To illustrate these notations, consider the following running example.

\paragraph{Running Example.}

\begin{figure}[ht]
    \centering
    \caption{Example Network Structure and Distance Matrix}
    \label{fig:toy}
    % Left Panel (a)
    \begin{subfigure}{0.44\textwidth}
        \centering
        \vspace{5pt} % Align baseline
        \begin{tikzpicture}[scale=1.5, node distance=1cm]
            % Nodes
            \node (i1) at (3.8,-0.8) {$i_1$};
            \node (i2) at (5.2,-0.8) {$i_2$};
            \node (i3) at (5.2,-2.2) {$i_3$};
            \node (i4) at (3.8,-2.2) {$i_4$};
            
            % Edges
            \draw[black, ultra thick] (4,-1) -- (5,-1);
            \draw[black, ultra thick] (5,-2) -- (4,-2);
        \end{tikzpicture}
        \caption{Network Structure}
        \label{fig:toy_a}
    \end{subfigure}
    \hfill
    % Right Panel (b)
    \begin{subfigure}{0.44\textwidth}
        \centering
        \vspace{0pt} % Align baseline
        $
        G = \begin{pmatrix} 
        0 & 1 & 2 & 2\\ 
        1 & 0 & 2 & 2\\
        2 & 2 & 0 & 1\\
        2 & 2 & 1 & 0
        \end{pmatrix}
        $
        \caption{Distance Matrix}
        \label{fig:toy_b}
    \end{subfigure}
    \notes{\textbf{Notes:} Panel (a) shows the network structure of the four units, and panel (b) displays the corresponding proximity (distance) matrix $G$.}
\end{figure}
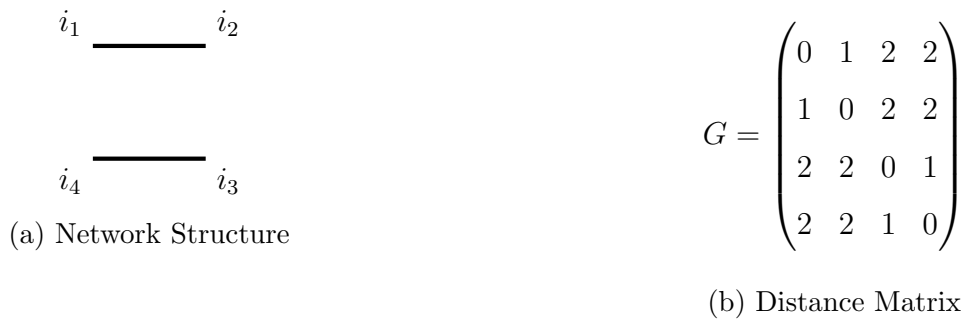

Consider four street segments, labeled $i_1$, $i_2$, $i_3$, and $i_4$, where two segments are considered adjacent if they are directly connected, as depicted in Figure~\ref{fig:toy}. Units $i_1$ and $i_2$ are connected, forming one area, while units $i_3$ and $i_4$ are connected, forming another area. For simplicity, I set the distance between units within the same area to $1$. In practice, distance measures should be determined by economic intuition, and the distance between units in different areas could be arbitrarily large. However, for the sake of this example, I set it to $2$.

Suppose the outcome of interest, $Y$, is the total number of crimes recorded over a year. In this example, exactly one unit receives a randomly assigned treatment to increase policing, so $P(d) = 1/4$ for each possible assignment. Let the observed treatment vector be $D^{obs} = (1,0,0,0)$, and the observed outcomes be $Y^{obs} = (2,4,3,2)$.

Table~\ref{tab:pon_ill} illustrates the potential outcome schedule under the design-based framework for all assignments that have positive probability. The first row corresponds to the observed dataset. Although all potential outcomes are fixed values, only those under the observed treatment assignment are actually observed. In general, since potential outcomes can depend on the treatment assignments across all units, there could theoretically be up to $2^N$ potential outcomes.

\begin{table}[ht]
\centering
\caption{Potential Outcome Schedule in the Example} 
\label{tab:pon_ill}
\setlength\tabcolsep{10pt}
\renewcommand{\arraystretch}{0.935}
{\small \begin{tabular}{c|c|c|c|c}
\hline \hline
Assignment $ D $ & \multicolumn{4}{c}{Potential Outcome $ Y_i $}   \\ \hline
& $ i_1 $ & $ i_2 $ & $ i_3 $ & $ i_4 $ \\
\hline 
\rowcolor{gray} 
$ (1,0,0,0) $ &  2  &  4  & 3 & 2  \\ 
\hline
$ (0,1,0,0) $  & \color{red}{?}  & \color{red}{?}  & \color{red}{?} & \color{red}{?}  \\
\hline 
$ (0,0,1,0) $  &  \color{red}{?} &  \color{red}{?} & \color{red}{?} & \color{red}{?} \\ 
\hline
$ (0,0,0,1) $ &   \color{red}{?} & \color{red}{?} & \color{red}{?} & \color{red}{?} \\
\hline \hline
\end{tabular}} \\
\notes{ \textbf{Notes:} The table shows the potential outcome schedule under the design-based view. The first row represents the observed assignment $ D^{obs} $, while potential outcomes denoted by \color{red}{?} \color{black} are unobserved values.}
\end{table}

%which is impractical to handle in practice. Focusing on how interference varies with distance motivates using an \textit{exposure mapping on distance} to help define the null hypothesis of interest and regularize the potential outcome schedule.

%Consider $\mathcal{P}$ as a known probability measure over $\{0,1\}^N$. The assignment vector is sampled with probability $P(d)=\mathcal{P}(D=d)=\mathcal{P}(\{d\})$ (\textit{use on P}), which is determined by the experimental design.

\setcounter{example}{0}
\subsection{Partially Sharp Null Hypothesis} \label{sec:null}

The term ``partially sharp null'' was first introduced by \citet{zhang2023}. I begin by providing a formal definition of the partially sharp null hypothesis.

\begin{definition}[Partially sharp null hypothesis] \label{def:partial}
A partially sharp null hypothesis holds if there exists a collection of subsets $ \{\mathcal{D}_i\}_{i=1}^N $, where each $ \mathcal{D}_i \subset \{0,1\}^N $, such that
\[
H_0: Y_i(d) = Y_i(d') \text{ for all } i \in \{1, \dots, N\}, \text{ and any } d, d' \in \mathcal{D}_i.
\]
\end{definition}

The partially sharp null hypothesis reduces dimensionality by restricting potential outcomes to vary only across certain subsets of assignments. The set $ \mathcal{D}_i $ can vary across units and is always a strict subset of $ \{0,1\}^N $, thus offering greater flexibility than the sharp null hypothesis, which corresponds to the case where $ \mathcal{D}_i = \{0,1\}^N $ for all $ i $. For instance, researchers can specify $ \mathcal{D}_i $ based on an exposure mapping---a function linking treatment assignments to exposure levels---to test outcome constancy within each exposure level, especially when concerned about potential misspecification \citep{hoshino2023}.

More generally, researchers can define alternative forms of $ \mathcal{D}_i $ that reflect specific hypotheses and research contexts, including cases where the null hypothesis is expressed as the intersection of multiple $ \mathcal{D}_i $ sets \citep{owusu2023, Puelz2022}. Appendix~\ref{app:inter} discusses extensions of the current framework to these more general and complex hypotheses. Although the method introduced in this paper is applicable to any partially sharp null hypothesis, I specifically focus on cases where $ \mathcal{D}_i $ is defined based on a distance measure.

\begin{definition}[Distance interval assignment set] \label{def:distance_set}
For a unit $ i \in \{1, \dots, N\} $ and a given distance $ \epsilon_s $, the distance interval assignment set is defined as
\[
\mathcal{D}_i (\epsilon_s) \equiv \left\{ d \in \{0,1\}^N : \sum_{j = 1}^N 1\{G_{i,j} \leq \epsilon_s\} d_j = 0 \right\}.
\]
When $ d \in \mathcal{D}_i (\epsilon_s) $, unit $ i $ is said to be in the distance interval $ (\epsilon_s, \infty) $.
\end{definition}

This definition involves two key concepts: $ \mathcal{D}_i (\epsilon_s) $ and the interval $ (\epsilon_s, \infty) $, both of which are specific to unit $ i $. The distance interval assignment set $ \mathcal{D}_i (\epsilon_s) $ maps a distance $ \epsilon_s $ to a set of treatment assignments where unit $ i $ is at least a distance $ \epsilon_s $ away from any treated units. For any $ \epsilon_s \geq 0 $, since $ G_{i,i} = 0 $, it follows that $ 1\{G_{i,i} \leq \epsilon_s\} = 1 $, implying that unit $ i $ is untreated ($ d_i = 0 $) for any assignment $ d \in \mathcal{D}_i(\epsilon_s) $. Specifically, when $ \epsilon_s = 0 $, all $ G_{i,j} $ for $ i \neq j $ are positive, which ensures that $ 1\{G_{i,j} \leq \epsilon_s\} = 0 $. As a result, there is no restriction on the treatment status of other units $ d_j $, and $ \mathcal{D}_i(0) $ includes all treatment assignments $ d $ where $ d_i = 0 $, while allowing others to be treated.\footnote{For any $ \epsilon_s < 0 $, since $ G_{i,j} \geq 0 $ for all $ i,j $, we have $ 1\{G_{i,j} \leq \epsilon_s\} = 0 $, meaning that $ \mathcal{D}_i (\epsilon_s) = \{0,1\}^N $, where all treatment assignments are included.}

The distance interval assignment set $ \mathcal{D}_i(a)/\mathcal{D}_i(b) $ corresponds to treatment assignments where unit $ i $ is within the distance interval $ (a,b] $. For any treatment assignment $ d $, the set $ \{i: d \in \mathcal{D}_i(a)/\mathcal{D}_i(b)\} $ contains all units that fall within the distance interval $ (a, b] $ relative to treated units.

Using the concept of distance interval assignment sets, I now define the partially sharp null hypothesis of interference based on distance.

\begin{definition}[Partially sharp null hypothesis of interference on distance $ \epsilon_s \geq 0 $] \label{def:gen}
The partially sharp null hypothesis of interference on distance $ \epsilon_s \geq 0 $ is defined as
\[
H^{\epsilon_s}_0: Y_i(d) = Y_i(d') \text{ for all } i \in \{1, \dots, N\}, \text{ and any } d, d' \in \mathcal{D}_i(\epsilon_s).
\]
\end{definition}

Under $ \mathcal{D}_i(\epsilon_s) $, all units within $ \epsilon_s $ distance of unit $ i $, as well as unit $ i $ itself, are not treated. Hence, this hypothesis asserts that no interference occurs beyond distance $ \epsilon_s $, meaning the potential outcomes for unit $ i $ remain unchanged for any treatment assignment where unit $ i $ is at least a distance $ \epsilon_s $ away from all treated units. Under this null hypothesis, the potential outcomes for unit $ i $ can be imputed for treatment assignment vectors that satisfy this distance condition, allowing for a partial imputation of outcomes. The interpretation of distance here is context-specific and depends on the nature of the interference in the particular application.\footnote{The paper focuses on constant distance across untreated units, but the same procedure can be implemented when allowing unit-level distance and a joint test for interference for both treated units and untreated units.}

\begin{example}[Spatial distance continued]  
In a setting where units represent street segments, for a given spatial distance $ \epsilon_s $ (e.g., 500 meters), $ \mathcal{D}_i (\epsilon_s) $ consists of all treatment assignments where unit $ i $ is at least 500 meters away from any treated street segments. The partially sharp null hypothesis $ H^{\epsilon_s}_0 $ tests whether spillover effects occur on an untreated unit located 500 meters away from any treated units.
\end{example}

\begin{example}[Network distance continued]  
Consider two schools, each with 100 students, where the goal is to test for cluster interference within schools. We assume that students within the same school are 100 units apart from each other and are infinitely distant from students in the other school. Setting $ \epsilon_s = 0 $, we test for interference within schools. Cluster interference is present if students' outcomes are affected by treatment assignments in their own school but not in the other school.\footnote{Setting $ \epsilon_s = 101 $ would test for interference across schools, but such a test may lack power in practice, as noted by \citet{Puelz2022}.}
\end{example}

\begin{example}[Link intensity continued]  
Consider a scenario where units represent individuals with cell phones, and the intensity of their connection is measured by the number of text messages exchanged, with a maximum of 50 messages per week. We define the ``distance'' between two individuals as $ 50 $ minus the number of messages exchanged. For $ \epsilon_s = 40 $, $ \mathcal{D}_i (\epsilon_s) $ represents all treatment assignments where unit $ i $ has exchanged fewer than 10 messages with any treated units. The partially sharp null hypothesis $ H^{\epsilon_s}_0 $ tests whether interference occurs for an untreated unit that has exchanged fewer than 10 messages with treated units.
\end{example}

The null hypothesis defined in Definition~\ref{def:gen} facilitates the assessment of whether, and to what extent, interference is present within a network. In many empirical settings, researchers seek to determine whether interference effects extend beyond a particular distance $ \epsilon_s $. When $ \epsilon_s > 0 $, this framework can be leveraged to delineate the neighborhood within which interference is operative, or to identify an appropriate comparison group for subsequent estimation.

\paragraph{Comparison to the Traditional t-Test.}
The traditional t-test compares units situated at various distances from treated units, but it is subject to two fundamental limitations. First, the distances between units and treated units are not random, even under randomized treatment assignment, which can induce bias absent further assumptions \citep{aronow2012, pollmann2023causal}. Second, standard large-sample approximations are complicated by the presence of complex clustering patterns \citep{morgan2021, blattman2021}. By contrast, the partially sharp null hypothesis articulated in Definition~\ref{def:gen} evaluates potential outcomes for the same unit under all treatment assignments in which it is farther than $ \epsilon_s $ from any treated unit. This approach requires only random assignment of treatment and circumvents biases that may arise from comparing outcomes across potentially non-comparable units.

\paragraph{Running Example Continued.} 

Suppose researchers wish to test for the existence of spillover effects using the partially sharp null hypothesis from Definition~\ref{def:gen} with $ \epsilon_s = 0 $:
\[
H^{0}_0: Y_i(d) = Y_i(d^\prime) \text{ for all } i \in \{1, \dots , N\}, \text{ and any } d, d^\prime \in \{0,1\}^N \text{ such that } d_i = d^\prime_i = 0.
\]

Throughout the paper, I use the above $ H^0_0 $ for illustration in the running example. This hypothesis implies that the potential outcome for any untreated unit $ i $ remains unchanged regardless of the treatment assignments of other units. The potential outcome schedule under $ H^0_0 $ is displayed in Table~\ref{tab:pon_toy}.

As shown in Table~\ref{tab:pon_toy}, the null hypothesis $ H^0_0 $ allows us to impute many of the previously missing potential outcomes. For example, since we observe the outcome when unit $ i_2 $ is not treated, we can impute other outcomes as long as $ i_2 $ remains untreated. Consequently, the outcome for $ i_2 $ when either unit $ i_3 $ or $ i_4 $ is treated is also 4.

\begin{table}[ht]
\centering 
\caption{Potential Outcome Schedule Under Partially Sharp Null $H^{0}_0$} 
\label{tab:pon_toy}
\setlength\tabcolsep{10pt}
\renewcommand{\arraystretch}{0.935}
{\small 
\begin{tabular}{c|c|c|c|c}
\hline \hline
Assignment $D$ & \multicolumn{4}{c}{Potential Outcome $Y_i$} \\ 
\hline
& $i_1$ & $i_2$ & $i_3$ & $i_4$ \\ 
\hline \rowcolor{gray}
$(1,0,0,0)$ & 2 & 4 & 3 & 2 \\ 
\hline
$(0,1,0,0)$ & \color{red}{?} & \color{red}{?} & 3 & 2 \\ 
\hline 
$(0,0,1,0)$ & \color{red}{?} & 4 & \color{red}{?} & 2 \\ 
\hline
$(0,0,0,1)$ & \color{red}{?} & 4 & 3 & \color{red}{?} \\ 
\hline \hline
\end{tabular}} \\
\notes{ \textbf{Notes:} The table shows the potential outcome schedule with the partially sharp null hypothesis under Definition \ref{def:gen} for the toy example. Assignment $D$ includes all potential assignments, with the first row representing the observed assignment $D^{obs}$. Potential outcomes marked in \color{red}{?} \color{black} are non-imputable values under the partially sharp null.}
\end{table}

However, as illustrated by Table \ref{tab:pon_toy}, the potential outcome schedule under $H^{0}_0$ still contains missing values, complicating the use of FRT \citep{zhang2023}. This highlights two key technical challenges for implementing randomization tests in the presence of interference.

\subsection{Two Technical Challenges for Randomization Tests} \label{sec:challenge}

\paragraph{Traditional Test Statistics.}

In practice, researchers often specify a distance $\epsilon_c$ to calibrate test power when interference diminishes with distance to treated units. For instance, in a spatial setting, $\epsilon_c$ is often set to a distance beyond which interference is considered negligible; for cluster interference, $\epsilon_c$ may be chosen to exceed the maximum distance within a cluster, so that no interference is expected across clusters. The purpose of this threshold is to separate units likely to be impacted by interference from those that can serve as clean controls.

A natural test statistic compares units within the distance interval $(\epsilon_s, \epsilon_c]$ to the treated group, while using units in the interval $(\epsilon_c, \infty)$ as a pure control group. When researchers lack prior knowledge to specify $\epsilon_c$, Appendix~\ref{sec:pure} proposes a sequential testing procedure to help select an appropriate threshold. Even if $\epsilon_c$ is misspecified and does not provide a perfectly clean control group, the proposed procedure remains valid, though it may reduce test power \citep{basse2024}.

For example, consider the difference-in-means estimator using the control distance $\epsilon_c$:
\[
T(Y(D^{obs}), D) =
\underbrace{\bar{Y}(D^{obs})_{\{i: D \in \mathcal{D}_i(\epsilon_s) \setminus \mathcal{D}_i(\epsilon_c)\}}}_{\text{Mean of neighbor group}}
-
\underbrace{\bar{Y}(D^{obs})_{\{i: D \in \mathcal{D}_i(\epsilon_c)\}}}_{\text{Mean of control group}},
\]
where for sets $A_i \subset \{0,1\}^N$, we define
\[
\bar{Y}(D^{obs})_{\{i: D \in A_i\}} =
\frac{\sum_{i=1}^N 1\{D \in A_i\} Y_{i}(D^{obs})}{\sum_{i=1}^N 1\{D \in A_i\}}.
\]
In particular, $A_i = \mathcal{D}_i(\epsilon_s) \setminus \mathcal{D}_i(\epsilon_c)$ corresponds to the distance interval $(\epsilon_s,\,\epsilon_c]$, while $A_i = \mathcal{D}_i(\epsilon_c)$ corresponds to $(\epsilon_c,\,\infty)$. The difference-in-means estimator is widely used in the literature (see, e.g., \citealp{basse2019b}; \citealp{Puelz2022}).

\paragraph{Running Example Continued.}

\begin{figure}[ht]
    \centering
    \caption{Example Network Structure with Treated, Neighbor, and Control Units}
    \label{fig:distance}
    % Panel (a)
    \begin{subfigure}{0.22\textwidth}
        \centering
        \vspace{0.2cm}
        \begin{tikzpicture}[node distance=1cm]
            % Nodes
            \node[circle, draw=green!60, color=red] (i1) at (3.8,-0.8) {$i_1$};
            \node[color=blue] (i2) at (5.2,-0.8) {$i_2$};
            \node[color=brown] (i3) at (5.2,-2.2) {$i_3$};
            \node[color=brown] (i4) at (3.8,-2.2) {$i_4$};
            
            % Edges
            \draw[black, ultra thick] (4,-1) -- (5,-1);
            \draw[black, ultra thick] (5,-2) -- (4,-2);
        \end{tikzpicture}
        \caption{Treated Unit: $i_1$}
    \end{subfigure}
    \hfill
    % Panel (b)
    \begin{subfigure}{0.22\textwidth}
        \centering
        \vspace{0.2cm}
        \begin{tikzpicture}[node distance=1cm]
            % Nodes
            \node[color=blue] (i1) at (3.8,-0.8) {$i_1$};
            \node[circle, draw=green!60, color=red] (i2) at (5.2,-0.8) {$i_2$};
            \node[color=brown] (i3) at (5.2,-2.2) {$i_3$};
            \node[color=brown] (i4) at (3.8,-2.2) {$i_4$};
            
            % Edges
            \draw[black, ultra thick] (4,-1) -- (5,-1);
            \draw[black, ultra thick] (5,-2) -- (4,-2);
        \end{tikzpicture}
        \caption{Treated Unit: $i_2$}
    \end{subfigure}
    \hfill
    % Panel (c)
    \begin{subfigure}{0.22\textwidth}
        \centering
        \vspace{0.2cm}
        \begin{tikzpicture}[node distance=1cm]
            % Nodes
            \node[color=brown] (i1) at (3.8,-0.8) {$i_1$};
            \node[color=brown] (i2) at (5.2,-0.8) {$i_2$};
            \node[circle, draw=green!60, color=red] (i3) at (5.2,-2.2) {$i_3$};
            \node[color=blue] (i4) at (3.8,-2.2) {$i_4$};
            
            % Edges
            \draw[black, ultra thick] (4,-1) -- (5,-1);
            \draw[black, ultra thick] (5,-2) -- (4,-2);
        \end{tikzpicture}
        \caption{Treated Unit: $i_3$}
    \end{subfigure}
    \hfill
    % Panel (d)
    \begin{subfigure}{0.22\textwidth}
        \centering
        \vspace{0.2cm}
        \begin{tikzpicture}[node distance=1cm]
            % Nodes
            \node[color=brown] (i1) at (3.8,-0.8) {$i_1$};
            \node[color=brown] (i2) at (5.2,-0.8) {$i_2$};
            \node[color=blue] (i3) at (5.2,-2.2) {$i_3$};
            \node[circle, draw=green!60, color=red] (i4) at (3.8,-2.2) {$i_4$};
            
            % Edges
            \draw[black, ultra thick] (4,-1) -- (5,-1);
            \draw[black, ultra thick] (5,-2) -- (4,-2);
        \end{tikzpicture}
        \caption{Treated Unit: $i_4$}
    \end{subfigure}

    \vspace{0.5em}
    \notes{\textbf{Notes:} Units with red circles are treated, units in blue are neighbors in the interval $(0, 1]$, and units in brown are control units in the interval $(1, \infty)$.}
\end{figure}
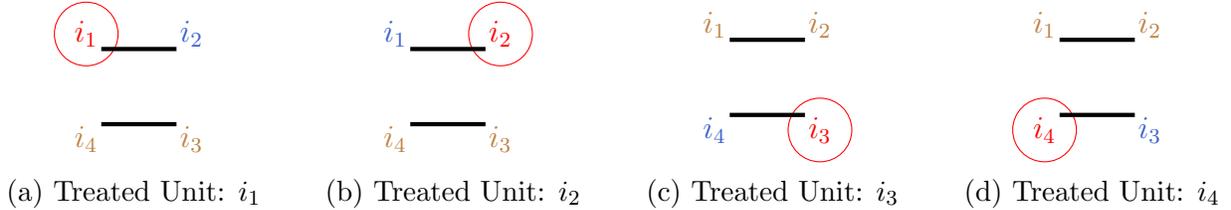

For the remainder of the running example, I set $\epsilon_c = 1$. Thus, there are two relevant distance intervals for the difference-in-means estimator: $(0, 1]$ and $(1, \infty)$. Figure~\ref{fig:distance} illustrates how these intervals change with different treatment assignments.

Applying traditional test statistics, such as the difference-in-means estimator, can be problematic when some potential outcomes remain unknown under $H^{0}_0$. Although the first row can be computed as $4-(3+2)/2=1.5$, Table~\ref{tab:trad_test_stat} shows that test statistics under non-observed treatment assignments still involve missing values. This occurs because randomization requires knowledge of all $ Y_i(d) $ values for the relevant assignment. This renders FRT inapplicable under the partially sharp null hypothesis and highlights two specific challenges that persist in more general settings:

\begin{table}[ht]
\centering 
\caption{Traditional Test Statistics Under Partially Sharp Null $H^{0}_0$} 
\label{tab:trad_test_stat}
\setlength\tabcolsep{10pt}
\renewcommand{\arraystretch}{0.935}
{\small
\begin{tabular}{c|c|c|c|c|c}
\hline \hline
Assignment $D$ & \multicolumn{4}{c|}{Potential Outcome $Y_i$} & $T(Y(D^{obs}), D)$ \\ \hline
& $i_1$ & $i_2$ & $i_3$ & $i_4$ &  \\
\hline \rowcolor{gray}
$(1,0,0,0)$ & 2 & \cellcolor{blue} 4 & \cellcolor{brown} 3 & \cellcolor{brown} 2 & 1.5 \\ 
\hline
$(0,1,0,0)$ & \cellcolor{blue} \color{red}{?} & \color{red}{?} & \cellcolor{brown} 3 & \cellcolor{brown} 2 & \color{red}{?} \\
\hline 
$(0,0,1,0)$ & \cellcolor{brown} \color{red}{?} & \cellcolor{brown} 4 & \color{red}{?} & \cellcolor{blue} 2 & \color{red}{?} \\ 
\hline
$(0,0,0,1)$ & \cellcolor{brown} \color{red}{?} & \cellcolor{brown} 4 & \cellcolor{blue} 3 & \color{red}{?} & \color{red}{?} \\
\hline \hline
\end{tabular}
}\\
\notes{ \textbf{Notes:} The table shows the potential outcome schedule under the partially sharp null hypothesis for the example. Assignment $D$ includes all potential assignments, with the first row representing the observed assignment $D^{obs}$. Potential outcomes marked in red question marks are non-imputable under the partially sharp null.}
\end{table}

First, only a subset of potential outcomes can be observed or imputed. For example, under $ H^{0}_0 $, if unit $ i_2 $ is treated, the hypothesis provides no information about the potential outcomes of unit $ i_1 $, leaving the potential outcomes for both $ i_1 $ and $ i_2 $ missing.

Second, the set of units with imputable outcomes depends on the treatment assignment. For instance, if unit $ i_3 $ is treated instead, the missing values now belong to $ i_1 $ and $ i_3 $, differing from other assignments.

The remainder of the paper develops new methods to address these challenges and enable valid inference in the presence of interference.

\section{Pairwise Imputation-based Randomization Test (PIRT)} \label{sec:PIRT}

For simplicity, I begin by fixing $\epsilon_s$ and $\epsilon_c$, deferring discussion of their selection until the end of this section. For each treatment assignment $d$, I formally define the set of units imputable under $H^{\epsilon_s}_0$.

\begin{definition}[Imputable units]\label{def:imput_units}
Given a treatment assignment $d \in \{0,1\}^N$ and a partially sharp null hypothesis $H^{\epsilon_s}_0$,
\[
\mathbb{I}(d) \equiv \{i \in \{1, \dots, N\}: d \in \mathcal{D}_i(\epsilon_s)\} \subseteq \{1, \dots, N\}
\]
is called the set of imputable units under treatment assignment $d$.
\end{definition}

The set of imputable units is the subset of units for which imputation is possible, corresponding to those in the distance interval $(\epsilon_s, \infty)$ under the partially sharp null hypothesis $H^{\epsilon_s}_0$. This concept shares a similar spirit with the ``super focal units'' in \citet{owusu2023}: given the observed treatment $D^{obs}$, the set $\mathbb{I}(D^{obs})$ includes all units with an imputable observed outcome. Units outside this set provide no additional information because their observed outcomes cannot be imputed to other treatment assignments under the partially sharp null. For example, if $\epsilon_s = 0$, then $\mathcal{D}_i(\epsilon_s)$ includes all assignments $d$ where $d_i = 0$, meaning $\mathbb{I}(D^{obs})$ consists of all units not treated under $D^{obs}$.

\paragraph{Running Example Continued.}

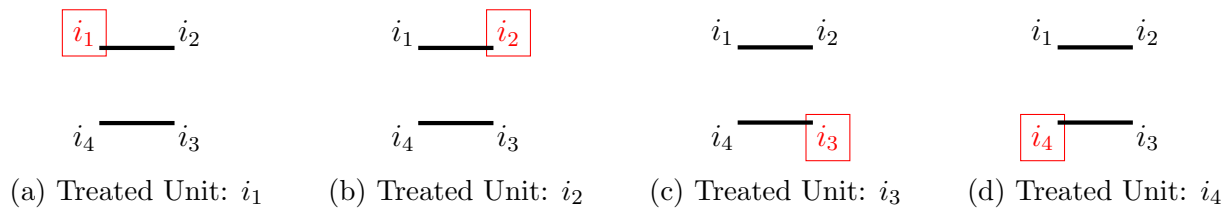
\begin{figure}[ht]
    \centering
    \caption{Example Network Structure with Imputable Units}
    \label{fig:imput}
    % Panel (a)
    \begin{subfigure}{0.22\textwidth}
        \centering
        \vspace{0.2cm}
        \begin{tikzpicture}[node distance=1cm]
            % Nodes
            \node[rectangle, draw=green!60, color=red] (i1) at (3.8,-0.8) {$i_1$};
            \node (i2) at (5.2,-0.8) {$i_2$};
            \node (i3) at (5.2,-2.2) {$i_3$};
            \node (i4) at (3.8,-2.2) {$i_4$};
            
            % Edges
            \draw[black, ultra thick] (4,-1) -- (5,-1);
            \draw[black, ultra thick] (5,-2) -- (4,-2);
        \end{tikzpicture}
        \caption{Treated Unit: $i_1$}
    \end{subfigure}
    \hfill
    % Panel (b)
    \begin{subfigure}{0.22\textwidth}
        \centering
        \vspace{0.2cm}
        \begin{tikzpicture}[node distance=1cm]
            % Nodes
            \node (i1) at (3.8,-0.8) {$i_1$};
            \node[rectangle, draw=green!60, color=red] (i2) at (5.2,-0.8) {$i_2$};
            \node (i3) at (5.2,-2.2) {$i_3$};
            \node (i4) at (3.8,-2.2) {$i_4$};
            
            % Edges
            \draw[black, ultra thick] (4,-1) -- (5,-1);
            \draw[black, ultra thick] (5,-2) -- (4,-2);
        \end{tikzpicture}
        \caption{Treated Unit: $i_2$}
    \end{subfigure}
    \hfill
    % Panel (c)
    \begin{subfigure}{0.22\textwidth}
        \centering
        \vspace{0.2cm}
        \begin{tikzpicture}[node distance=1cm]
            % Nodes
            \node (i1) at (3.8,-0.8) {$i_1$};
            \node (i2) at (5.2,-0.8) {$i_2$};
            \node[rectangle, draw=green!60, color=red] (i3) at (5.2,-2.2) {$i_3$};
            \node (i4) at (3.8,-2.2) {$i_4$};
            
            % Edges
            \draw[black, ultra thick] (4,-1) -- (5,-1);
            \draw[black, ultra thick] (5,-2) -- (4,-2);
        \end{tikzpicture}
        \caption{Treated Unit: $i_3$}
    \end{subfigure}
    \hfill
    % Panel (d)
    \begin{subfigure}{0.22\textwidth}
        \centering
        \vspace{0.2cm}
        \begin{tikzpicture}[node distance=1cm]
            % Nodes
            \node (i1) at (3.8,-0.8) {$i_1$};
            \node (i2) at (5.2,-0.8) {$i_2$};
            \node (i3) at (5.2,-2.2) {$i_3$};
            \node[rectangle, draw=green!60, color=red] (i4) at (3.8,-2.2) {$i_4$};
            
            % Edges
            \draw[black, ultra thick] (4,-1) -- (5,-1);
            \draw[black, ultra thick] (5,-2) -- (4,-2);
        \end{tikzpicture}
        \caption{Treated Unit: $i_4$}
    \end{subfigure}
    \notes{\textbf{Notes:} Treated units are marked with red rectangles, while imputable units are shown in black.} 
\end{figure}

Under $H^0_0$, as illustrated in Figure~\ref{fig:imput}, when unit $i_1$ is treated, units $i_2$ to $i_4$ belong to the imputable set. Similarly, when unit $i_2$ is treated, units $i_1$, $i_3$, and $i_4$ are imputable. This setting corresponds to a special case where all untreated units are imputable. In more general settings, however, the composition of the imputable set depends on the value of $\epsilon_s$ specified in the null hypothesis.

As demonstrated in Figure~\ref{fig:imput}, in general $\mathbb{I}(d) \neq \mathbb{I}(d')$ for different assignments $d$ and $d'$. For instance, when testing for spillover effects among friends in a network, the set of friends whose outcomes are imputable will change across treatment assignments, reflecting the underlying social structure and the value of $\epsilon_s$.

In practice, $\mathbb{I}(D^{obs})$ could sometimes be empty, depending on the network structure and the specific partially sharp null hypothesis. If no units meet the required criteria (i.e., \( \mathbb{I}(D^{obs}) \) is empty), one approach is to reject the null hypothesis \( \alpha \) percent of the time, in line with the desired significance level. This ensures control of the test's size, even in cases where the imputable set is empty. However, to achieve power in such cases, additional data or a different study design may be necessary. See Appendix \ref{app:empty} for further discussion.

The set of imputable units can also be defined under the sharp null hypothesis, though in this case, $\mathbb{I}(d) = \{1, \dots, N\}$ for any assignment $d$, meaning all units are imputable under the sharp null. Therefore, there has been less focus on the imputable units set in the randomization tests literature. 

\subsection{Pairwise Imputable Statistics} \label{sec:test_stat}

To address the first technical challenge—the presence of missing potential outcomes—I construct a class of test statistics that remain valid by relying only on the observed, imputable outcomes. 

\begin{definition}[Pairwise Imputable Statistic]\label{def:test_stat}
Let
$
T: \mathbb{R}^N \times \{0,1\}^N \times \{0,1\}^N \to \mathbb{R} \cup \{\infty\}
$
be a measurable function. We say that $T$ is \emph{pairwise imputable} if, for any $d, d' \in \{0,1\}^N$ and any pair of outcome vectors $Y, Y' \in \mathbb{R}^N$, the following holds:
\[
\text{If } Y_i = Y'_i \text{ for all } i \in \mathbb{I}(d) \cap \mathbb{I}(d'), \text{ then } T(Y, d, d') = T(Y', d, d').
\]
That is, $T$ depends only on the entries of $Y$ indexed by the intersection $\mathbb{I}(d) \cap \mathbb{I}(d')$.
\end{definition}

The set $\mathbb{I}(d) \cap \mathbb{I}(d')$ in Definition~\ref{def:test_stat} closely parallels the set $H$ in Definition~1 of \citet{zhang2023}, capturing units whose potential outcomes are jointly imputable under both assignments. At first glance, the pairwise imputability requirement may appear restrictive. However, it is flexible enough to encompass commonly used test statistics with slight modification. For example, the classic difference in means can be written as
\[
T(Y(D^{\mathrm{obs}}), D, D^{\mathrm{obs}}) = \underbrace{\bar{Y}_{\mathbb{I}(D^{\mathrm{obs}})}(D^{\mathrm{obs}})_{\{i: D \in \mathcal{D}_i(\epsilon_s) \setminus \mathcal{D}_i(\epsilon_c)\}}}_{\text{Mean of \textit{imputable} neighbor}} \\
 -\ \underbrace{\bar{Y}_{\mathbb{I}(D^{\mathrm{obs}})}(D^{\mathrm{obs}})_{\{i: D \in \mathcal{D}_i(\epsilon_c)\}}}_{\text{Mean of \textit{imputable} control}}.
\]
where, for any collection of sets $A_i \subset \{0,1\}^N$,
\[
\bar{Y}_{\mathbb{I}(D^{\mathrm{obs}})}(D^{\mathrm{obs}})_{\{i: D \in A_i\}} =
\frac{\sum_{i \in \mathbb{I}(D^{\mathrm{obs}})} 1\{D \in A_i\} Y_{i}(D^{\mathrm{obs}})}
     {\sum_{i \in \mathbb{I}(D^{\mathrm{obs}})} 1\{D \in A_i\}}\,.
\]

This construction generalizes the classical sharp null setting: when all potential outcomes are observed (i.e., under the sharp null hypothesis), $\mathbb{I}(d) \cap \mathbb{I}(d') = \{1, \ldots, N\}$ for all $d, d'$, and any standard test statistic---such as the difference in means---falls within this framework \citep{Imbens_Rubin_2015}. In particular, when all units are imputable (i.e., $\mathbb{I}(D^{\mathrm{obs}}) = \{1, \ldots, N\}$), the above reduces to the standard difference in means, with groupings determined by the treatment assignment $D$.

If, for a given assignment, no unit in $\mathbb{I}(D^{\mathrm{obs}})$ falls into one of the specified intervals, the corresponding sample mean is undefined. In such cases, I set $T = \max(Y^{\mathrm{obs}}) - \min(Y^{\mathrm{obs}})$, ensuring the test remains valid, albeit conservative.\footnote{Any constant exceeding the observed statistic would suffice.} In practice, this situation did not arise in my empirical application and is unlikely in bipartite experiments with moderate~$\epsilon_s$. For further discussion, see Appendix~\ref{app:empty}.\footnote{Alternatively, to maximize power, conditional randomization testing can be used to exclude problematic assignments \citep{zhang2023}.}

\paragraph{Running Example Continued.}
Consider the test statistic
\[
T(Y(D^{\mathrm{obs}}), D, D^{\mathrm{obs}}) = \bar{Y}_{\mathbb{I}(D^{\mathrm{obs}})}(D^{\mathrm{obs}})_{\{i: D \in \mathcal{D}_i(0)\setminus\mathcal{D}_i(1)\}} - \bar{Y}_{\mathbb{I}(D^{\mathrm{obs}})}(D^{\mathrm{obs}})_{\{i: D \in \mathcal{D}_i(1)\}}.
\]

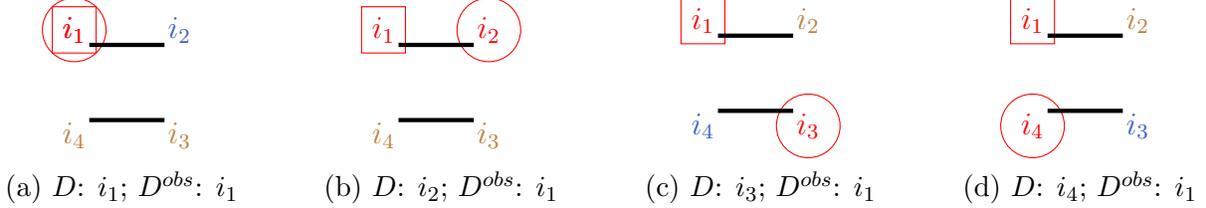
\begin{figure}[ht]
    \centering
    \caption{Illustration of Imputable Neighbor and Control Units for $T(Y(D^{\mathrm{obs}}), D, D^{\mathrm{obs}})$}
    \label{fig:stats}
    % Panel (a)
    \begin{subfigure}{0.22\textwidth}
        \centering
        \begin{tikzpicture}[node distance=1cm]
            \node[rectangle, draw=green!60, color=red] (i1) at (3.8,-0.8) {$i_1$};
            \node[circle, draw=green!60, color=red] (i1) at (3.8,-0.8) {$i_1$};
            \node[color=blue] (i2) at (5.2,-0.8) {$i_2$};
            \node[color=brown] (i3) at (5.2,-2.2) {$i_3$};
            \node[color=brown] (i4) at (3.8,-2.2) {$i_4$};
            % Edges
            \draw[black, ultra thick] (4,-1) -- (5,-1) (5,-2) -- (4,-2);
        \end{tikzpicture}
        \caption{$D$: $i_1$; $D^{\mathrm{obs}}$: $i_1$}
    \end{subfigure}
    \hfill
    % Panel (b)
    \begin{subfigure}{0.22\textwidth}
        \centering
        \begin{tikzpicture}[node distance=1cm]
            \node[rectangle, draw=green!60, color=red] (i1) at (3.8,-0.8) {$i_1$};
            \node[circle, draw=green!60, color=red] (i2) at (5.2,-0.8) {$i_2$};
            \node[color=brown] (i3) at (5.2,-2.2) {$i_3$};
            \node[color=brown] (i4) at (3.8,-2.2) {$i_4$};
            % Edges
            \draw[black, ultra thick] (4,-1) -- (5,-1) (5,-2) -- (4,-2);
        \end{tikzpicture}
        \caption{$D$: $i_2$; $D^{\mathrm{obs}}$: $i_1$}
    \end{subfigure}
    \hfill
    % Panel (c)
    \begin{subfigure}{0.22\textwidth}
        \centering
        \begin{tikzpicture}[node distance=1cm]
            \node[rectangle, draw=green!60, color=red] (i1) at (3.8,-0.8) {$i_1$};
            \node[color=brown] (i2) at (5.2,-0.8) {$i_2$};
            \node[circle, draw=green!60, color=red] (i3) at (5.2,-2.2) {$i_3$};
            \node[color=blue] (i4) at (3.8,-2.2) {$i_4$};
            % Edges
            \draw[black, ultra thick] (4,-1) -- (5,-1) (5,-2) -- (4,-2);
        \end{tikzpicture}
        \caption{$D$: $i_3$; $D^{\mathrm{obs}}$: $i_1$}
    \end{subfigure}
    \hfill
    % Panel (d)
    \begin{subfigure}{0.22\textwidth}
        \centering
        \begin{tikzpicture}[node distance=1cm]
            \node[rectangle, draw=green!60, color=red] (i1) at (3.8,-0.8) {$i_1$};
            \node[color=brown] (i2) at (5.2,-0.8) {$i_2$};
            \node[color=blue] (i3) at (5.2,-2.2) {$i_3$};
            \node[circle, draw=green!60, color=red] (i4) at (3.8,-2.2) {$i_4$};
            % Edges
            \draw[black, ultra thick] (4,-1) -- (5,-1) (5,-2) -- (4,-2);
        \end{tikzpicture}
        \caption{$D$: $i_4$; $D^{\mathrm{obs}}$: $i_1$}
    \end{subfigure}
    \notes{\textbf{Notes:} Red circles indicate treated units in $D$, which determine neighbor units in the interval $(0,1]$ and control units in $(1,\infty)$. Red rectangles indicate treated units in $D^{\mathrm{obs}}$, which determine imputable units.}
\end{figure}

Table~\ref{tab:pair_stat} presents the corresponding values for the first and second terms of the test statistic, while Figure~\ref{fig:stats} provides a visual representation of how we determine the imputable neighbor units and imputable control units.

\begin{table}[ht]
\centering
\caption{Constructing a Pairwise Imputable Statistic}
\label{tab:pair_stat}
\setlength\tabcolsep{8pt}
\renewcommand{\arraystretch}{0.935}
{\small
\begin{tabular}{c|>{\columncolor{red}}c|c|c|c|c|c|c}
\hline \hline
Assignment $D$ & \multicolumn{4}{c|}{Potential Outcome $Y_i$} & \multicolumn{2}{c|}{\scriptsize $\bar{Y}_{\mathbb{I}(D^{\mathrm{obs}})}(D^{\mathrm{obs}})$} & {\scriptsize $T(Y(D^{\mathrm{obs}}), D, D^{\mathrm{obs}})$} \\ \hline
& $i_1$ & $i_2$ & $i_3$ & $i_4$ & {\scriptsize ${\{i: D \in \mathcal{D}_i(0)\setminus\mathcal{D}_i(1)\}}$} & {\scriptsize ${\{i: D \in \mathcal{D}_i(1)\}}$} &  \\
\hline \rowcolor{gray}
$(1,0,0,0)$ & 2 & \cellcolor{blue} 4 & \cellcolor{brown} 3 & \cellcolor{brown} 2 & 4 & 2.5 & 1.5 \\
\hline
$(0,1,0,0)$ & \color{red}{?} & \color{red}{?} & \cellcolor{brown} 3 & \cellcolor{brown} 2 & \color{red}{?} & 2.5 & \color{red}{2} \\
\hline
$(0,0,1,0)$ & \color{red}{?} & \cellcolor{brown} 4 & \color{red}{?} & \cellcolor{blue} 2 & 2 & 4 & -2 \\
\hline
$(0,0,0,1)$ & \color{red}{?} & \cellcolor{brown} 4 & \cellcolor{blue} 3 & \color{red}{?} & 3 & 4 & -1 \\
\hline\hline
\end{tabular}
}
\\
\notes{
\textbf{Notes:} Assignment $D$ includes all potential assignments, with the first row corresponding to the observed assignment $D^{\mathrm{obs}}$. Potential Outcome $Y_i$ is the potential outcome of each unit under the null $H^0_0$, with red question marks representing missing values. Unit $i_1$ does not belong to set $\mathbb{I}(D^{\mathrm{obs}})$, so the entire column is marked in red. $\bar{Y}_{\mathbb{I}(D^{\mathrm{obs}})}(D^{\mathrm{obs}})$ with ${\{i: D \in \mathcal{D}_i(0)\setminus\mathcal{D}_i(1)\}}$ is the mean potential outcome for units in the distance interval $(0,1]$, marked in blue. $\bar{Y}_{\mathbb{I}(D^{\mathrm{obs}})}(D^{\mathrm{obs}})$ with ${\{i: D \in \mathcal{D}_i(1)\}}$ is the mean potential outcome for units in the distance interval $(1,\,\infty)$. $T = \max(Y^{\mathrm{obs}}) - \min(Y^{\mathrm{obs}})$, marked in red when one of the mean values is undefined.
}
\end{table}

As illustrated in Figure~\ref{fig:stats}, $D^{\mathrm{obs}}$ refers to the scenario where unit $i_1$ is treated, so the set of imputable units remains the same across different potential assignments $D$. However, the potential assignment $D$ itself can change, thereby altering which units belong to the neighborhood and control sets. When $D = D^{\mathrm{obs}}$ and unit $i_1$ is treated, the first term $\bar{Y}_{\mathbb{I}(D^{\mathrm{obs}})}(D^{\mathrm{obs}})_{\{i: D \in \mathcal{D}_i(0)\setminus\mathcal{D}_i(1)\}}$ corresponds to the outcome of $i_2$, while the second term $\bar{Y}_{\mathbb{I}(D^{\mathrm{obs}})}(D^{\mathrm{obs}})_{\{i: D \in \mathcal{D}_i(1)\}}$ is the mean outcome of $i_3$ and $i_4$. When unit $i_2$ is treated, there are no imputable units in the neighborhood set; in this case, we define $T = \max(Y^{\mathrm{obs}}) -\min(Y^{\mathrm{obs}}) = 2$ to ensure the test's validity.

The proposed test statistic satisfies Definition~\ref{def:test_stat} because only units in the intersection $\mathbb{I}(D^{\mathrm{obs}})\cap\mathbb{I}(D)$ are used to construct it. For example, if $D^{\mathrm{obs}}$ has $i_1$ treated and $D$ has $i_3$ treated, then $\mathbb{I}(D^{\mathrm{obs}}) = \{i_2, i_3, i_4\}$ and $\mathbb{I}(D) = \{i_1, i_2, i_4\}$, so their intersection is $\{i_2, i_4\}$. As shown in the third row of Table~\ref{tab:pair_stat}, the test statistic in this case depends only on the outcomes of units $i_2$ and $i_4$.

\begin{sloppypar}
Additionally, rank-based statistics can be incorporated by excluding non-imputable units and reranking the remaining units. Following \citet{Imbens_Rubin_2015}, define the rank of unit~$i$ among the imputable units as
\begin{align*}
R_i
&\equiv R_i\bigl(Y_{\mathbb{I}(D^{\mathrm{obs}}) \cap \mathbb{I}(D)}(D^{\mathrm{obs}})\bigr) \\
&= \sum_{j \in \mathbb{I}(D^{\mathrm{obs}}) \cap \mathbb{I}(D)} 1\{Y_j(D^{\mathrm{obs}}) < Y_i(D^{\mathrm{obs}})\} \\
&\quad + 0.5\Bigl(1
    + \sum_{j \in \mathbb{I}(D^{\mathrm{obs}}) \cap \mathbb{I}(D)} 1\{Y_j(D^{\mathrm{obs}}) = Y_i(D^{\mathrm{obs}})\}\Bigr) \\
&\quad - \frac{1 + \left|\mathbb{I}(D^{\mathrm{obs}}) \cap \mathbb{I}(D)\right|}{2}.
\end{align*}
The corresponding test statistic is then
$
T(Y(D^{\mathrm{obs}}), D, D^{\mathrm{obs}}) = \bar{R}_{\{i: D \in \mathcal{D}_i(\epsilon_s) \setminus \mathcal{D}_i(\epsilon_c)\}} - \bar{R}_{\{i: D \in \mathcal{D}_i(\epsilon_c)\}},
$
where $\bar{R}$ denotes the mean rank within the specified group.

If $Y_i(D^{\mathrm{obs}}) = Y_i(D)$ for all $i \in \mathbb{I}(D^{\mathrm{obs}}) \cap \mathbb{I}(D)$, then $R_i(Y_{\mathbb{I}(D^{\mathrm{obs}}) \cap \mathbb{I}(D)}(D^{\mathrm{obs}})) = R_i(Y_{\mathbb{I}(D^{\mathrm{obs}}) \cap \mathbb{I}(D)}(D))$; that is, the ranks remain unchanged. Therefore, $T(Y(D^{\mathrm{obs}}), D, D^{\mathrm{obs}}) = T(Y(D), D, D^{\mathrm{obs}})$, so the statistic satisfies Definition~\ref{def:test_stat}.

For further discussion of test statistic selection in randomization tests and network settings, see Section~5 of \citet{Imbens_Rubin_2015}, \citet{athey2018}, and \citet{hoshino2023}. All test statistics can also be adapted to their absolute-value forms for two-sided testing.
\end{sloppypar}

While the method remains valid without covariate adjustments, incorporating them may improve the test's power in practice \citep{ding2021}. See Appendix~\ref{app:covar} for a discussion on incorporating covariates. Moreover, since the proposed method is finite-sample valid, researchers can conduct subgroup analyses when different patterns of interference are expected across covariates.

Following Definition~\ref{def:test_stat} of pairwise imputable statistics, I can derive a property that allows calculation of test statistics using only observed information:

\begin{proposition}\label{prop:validity}
    Suppose the partially sharp null hypothesis $H^{\epsilon_s}_0$ is true. Suppose $T(Y(d), d, d^\prime)$ is a pairwise imputable statistic. Then,
    \[
    T(Y(d), d, d^\prime) = T(Y(d^\prime), d, d^\prime)
    \]
    for any $d, d^\prime \in \{0,1\}^N$.
\end{proposition}
The proof is provided in Appendix~\ref{app:proof}. By Proposition~\ref{prop:validity}, letting $d = D$ and $d^\prime = D^{\mathrm{obs}}$, we have $T(Y(D), D, D^{\mathrm{obs}}) = T(Y(D^{\mathrm{obs}}), D, D^{\mathrm{obs}})$ under the null $H^{\epsilon_s}_0$, ensuring the construction of the counterfactual test statistic.

\subsection{Unconditional Randomization Test}\label{sec:uncon}

In this paper, I focus on the unconditional randomization test framework, which is defined as follows:

\begin{definition}[Unconditional randomization test]\label{def:uncon_test}
An unconditional randomization test $ \phi: \{0,1\}^N \rightarrow [0,1] $ is defined such that for any $ D^{\mathrm{obs}} \in \{0,1\}^N $,
\[
\phi(D^{\mathrm{obs}}) = Q(\tilde{p}(D^{\mathrm{obs}}), \alpha),
\]
where $ Q: [0,1] \times [0,1] \rightarrow [0,1] $ is a measurable function, $ \alpha $ is the nominal significance level, and $ \tilde{p}(D^{\mathrm{obs}}) $ is given by
\[
\tilde{p}(D^{\mathrm{obs}}) = \sum_{d \in \{0,1\}^N} g(D^{\mathrm{obs}}, d)\, P(D = d),
\]
with $ P $ denoting the pre-specified probability distribution over treatment assignments, and $ g: \{0,1\}^N \times \{0,1\}^N \rightarrow \{0,1\} $ a measurable function.
\end{definition}

The key feature of the unconditional randomization test is that the probability of rejection, $\phi(D^{\mathrm{obs}})$, is computed by randomizing the treatment assignment according to the same probability distribution $P$ that governs the original assignment. This stands in contrast to methods in the existing literature, such as \citet{athey2018}, where the rejection function is based on randomizing the treatment assignment within a conditional probability space, conditioning on certain events. One example is the simple randomization test, which uses pairwise imputable statistics and constructs \textit{p}-values analogously to the classic Fisher Randomization Test (FRT).

\begin{definition}[Simple randomization test]\label{def:naive_test}
A \textit{simple randomization test} is an unconditional randomization test defined by $\phi(D^{\mathrm{obs}}) = 1\{pval(D^{\mathrm{obs}}) \leq \alpha\}$, where $pval(D^{\mathrm{obs}}): \{0,1\}^N \rightarrow [0,1]$ is the \textit{p}-value function given by 
\[
pval(D^{\mathrm{obs}}) = P(T(Y(D^{\mathrm{obs}}), D, D^{\mathrm{obs}}) \geq T(Y(D^{\mathrm{obs}}), D^{\mathrm{obs}}, D^{\mathrm{obs}})) \text{ for } D \sim P,
\]
and $T(Y, d, d')$ denotes a pairwise imputable statistic.
\end{definition}

\paragraph{Running Example Continued.}

Using the pairwise imputable statistics $T(Y(D^{\mathrm{obs}}), D, D^{\mathrm{obs}})$ and following Table~\ref{tab:pair_stat}, we can construct Table~\ref{tab:naive} with the test statistics for each assignment.

\begin{table}[ht]
\centering 
\caption{Simple Randomization Test in the Example} 
\label{tab:naive}
\setlength\tabcolsep{10pt}
\renewcommand{\arraystretch}{0.935}
{\small
\begin{tabular}{c|>{\columncolor{red}}c|c|c|c|c}
\hline \hline
Assignment $D$ & \multicolumn{4}{c|}{Potential Outcome $Y_i$} & {\scriptsize $T(Y(D^{\mathrm{obs}}), D, D^{\mathrm{obs}})$} \\ \hline
& $i_1$ & $i_2$ & $i_3$ & $i_4$ &   \\
\hline \rowcolor{gray}
$(1,0,0,0)$ &  2  & \cellcolor{blue} 4  & \cellcolor{brown} 3 &  \cellcolor{brown} 2 &  1.5 \\ 
\hline
$(0,1,0,0)$  &  \color{red}{?}  & \color{red}{?}  &  \cellcolor{brown} 3 & \cellcolor{brown} 2 & \color{red}{2}\\
\hline 
$(0,0,1,0)$  &   \color{red}{?} &  \cellcolor{brown} 4 & \color{red}{?} & \cellcolor{blue} 2 & -2 \\ 
\hline
$(0,0,0,1)$ &  \color{red}{?} & \cellcolor{brown}  4 & \cellcolor{blue} 3 & \color{red}{?} & -1 \\
\hline \hline
\end{tabular}
}\\
\notes{\textbf{Notes:} Assignment $D$ includes all possible assignments, with the first row representing the observed assignment $D^{\mathrm{obs}}$. Potential Outcome $Y_i$ denotes the potential outcome for each unit under the null $H_0^0$, while red question marks indicate missing values. Unit $i_1$ does not belong to the set $\mathbb{I}(D^{\mathrm{obs}})$, so its column is highlighted in red. Blue cells denote units used to calculate the mean value in the first term of the test statistic. $T(Y(D^{\mathrm{obs}}), D, D^{\mathrm{obs}})$ are the test statistics for each $D$, fixing $D^{\mathrm{obs}}$ where unit $i_1$ is treated.}
\end{table}

Based on Table~\ref{tab:naive} and following Definition~\ref{def:naive_test}, the \textit{p}-value is $2/4$. However, one might question whether this procedure guarantees finite-sample validity---specifically, whether it satisfies the condition $E_P(\phi(D^{\mathrm{obs}})) \leq \alpha$ under the null hypothesis.

\paragraph{Investigating Finite-Sample Validity.}

Although pairwise imputable statistics are used, naively constructing the \textit{p}-value as defined in the classic FRT does not guarantee the test's validity. For the test to be valid, the following condition must hold under the partially sharp null hypothesis:
\[
T(Y(D^{\mathrm{obs}}), D, D^{\mathrm{obs}})  \stackrel{d}{=} T(Y(D^{\mathrm{obs}}), D^{\mathrm{obs}}, D^{\mathrm{obs}}),
\]
where $ \stackrel{d}{=} $ indicates equality in distribution. The distribution on the left-hand side (LHS) is with respect to $ D $, while the distribution on the right-hand side (RHS) is with respect to $ D^{\mathrm{obs}} $.

By Proposition \ref{prop:validity}, under the null hypothesis, we also have:
\[
T(Y(D^{\mathrm{obs}}), D, D^{\mathrm{obs}}) \nulleq T(Y(D), D, D^{\mathrm{obs}}).
\]
\begin{sloppypar}
Here, $ \nulleq $ denotes equality under the null hypothesis. However, the term $ T(Y(D^{\mathrm{obs}}), D^{\mathrm{obs}}, D^{\mathrm{obs}}) $, being induced by the randomness of $ D^{\mathrm{obs}} $, satisfies:
\end{sloppypar}
\[
T(Y(D^{\mathrm{obs}}), D^{\mathrm{obs}}, D^{\mathrm{obs}}) \stackrel{d}{=} T(Y(D), D, D).
\]

Thus, for the test to maintain validity, we require:
\[
T(Y(D), D, D^{\mathrm{obs}}) \stackrel{d}{=} T(Y(D), D, D).
\]

This condition is not guaranteed under the partially sharp null hypothesis because $ \mathbb{I}(D^{\mathrm{obs}}) \neq \mathbb{I}(D) $ in general. Different treatment assignments $ D $ result in different sets of imputable units, leading to variability in $ \mathbb{I}(D) $. This is a key technical challenge. In the special case of testing the sharp null hypothesis, where $ \mathbb{I}(D^{\mathrm{obs}}) = \{1, \dots, N\} = \mathbb{I}(D) $, the validity trivially holds.

To address the challenges posed by varying imputable unit sets, previous literature proposes a remedy through the design of a conditioning event, consisting of a fixed subset of imputable units (referred to as \textit{focal units}) and a fixed subset of assignments (\textit{focal assignments}). CRTs are then performed by conducting FRTs within this conditioning event. While this approach has been influential, it also presents certain practical considerations.

First, as \citet{zhang2023} highlighted, there exists a trade-off between the sizes of focal units and focal assignments: expanding the subset of treatment assignments typically requires a reduction in the subset of experimental units. This trade-off may result in less information being utilized within the conditioning events, which can impact the power of the test. Second, constructing the conditioning event introduces an additional layer of computational complexity. This naturally raises the question of whether unconditional randomization testing remains valid in finite samples.

Whereas previous approaches ensure the validity of randomization testing by carefully designing a fixed subset of units, my method takes a different route. It avoids fixing the subset of units during implementation and instead achieves valid testing through a carefully constructed \textit{p}-value calculation, thereby ensuring finite-sample validity without relying on conditioning events.

\subsection{Pairwise Comparison-Based \textit{p}-values} \label{sec:pvalue_pair}
\begin{sloppypar}
Building on the selective inference literature \citep{wen2023residual, guan2023conformal}, the central idea is to compute \textit{p}-values by aggregating pairwise inequality comparisons between $T(Y(D^{obs}), d^r, D^{obs})$ and $T(Y(D^{obs}), D^{obs}, d^r)$. When the null hypothesis is false, $T(Y(D^{obs}), D^{obs}, d^r)$ remains relatively large across different $d^r$, since the distance interval for each unit is fixed by $D^{obs}$. The change in $d^r$ only alters the set of units used in the test statistic, and rejection of the null remains possible when the units in the neighborhood set tend to have high outcome values. Consequently, we would expect a small \textit{p}-value, as the probability that $T(Y(D^{obs}), d^r, D^{obs})$ exceeds $T(Y(D^{obs}), D^{obs}, d^r)$ is low.
\end{sloppypar}

Formally, I refer to any randomization test with \textit{p}-values constructed through this pairwise comparison method as a ``PIRT'' (Pairwise Imputation-based Randomization Test).

\begin{definition}[PIRT] \label{def:pvalue_pair}
The PIRT is an unconditional randomization test defined by
\[
    \phi^{pair}(D^{obs}) = 1\left\{ pval^{pair}(D^{obs}) \leq \alpha/2 \right\},
\]
where $pval^{pair}(D^{obs}): \{0,1\}^N \rightarrow [0,1]$ is the \textit{p}-value function given by
\[
    pval^{pair}(D^{obs}) = P\left( T(Y(D^{obs}), D, D^{obs}) \geq T(Y(D^{obs}), D^{obs}, D) \right), \quad \text{for } D \sim P,
\]
and $T(Y, d, d')$ denotes a pairwise imputable statistic.
\end{definition}

\begin{theorem} \label{theo:validity_pair}
Suppose the partially sharp null hypothesis $ H^{\epsilon_s}_0 $ holds. Then, the PIRT, as defined in Definition~\ref{def:pvalue_pair}, satisfies $ \mathbb{E}_P[\phi^{pair}(D^{obs})] < \alpha $ for any $ \alpha \in (0,1) $, where the expectation is taken with respect to $ D^{obs} \sim P $.
\end{theorem}

See Appendix~\ref{app:proof} for the proof. The validity result follows from Proposition~\ref{prop:validity} and the conformal lemma in the conformal prediction literature \citep{guan2023conformal}. Specifically, under the null, $T(Y(D^{obs}), D^{obs}, D) = T(Y(D), D^{obs}, D)$, which coincides with the randomized test statistic $T(Y(D^{obs}), D, D^{obs})$ when swapping the roles of $D$ and $D^{obs}$. Thus, the pairwise comparison is symmetric, which allows the application of the conformal lemma to complete the proof.

Theorem~\ref{theo:validity_pair} provides a worst-case validity guarantee, analogous to those in cross-conformal prediction and jackknife+ methods, due to certain pathological cases \citep{vovk2018a, barber2021predictive, guan2023conformal}. As in that literature, the test empirically achieves size control at level $\alpha/2$, as demonstrated in Section~\ref{sec:simu}, but this property cannot be established theoretically due to the existence of pathological examples.\footnote{See Appendix~\ref{app:min_PIRT} for a more conservative minimization-based PIRT, which achieves theoretical size control with a rejection threshold of $\alpha$.}

Even with a large sample, an unbiased estimator of the \textit{p}-value can be computed using Algorithm~\ref{algo:pair}, which calculates the \textit{p}-value as the average over $1 + R$ draws, where $r=0$ corresponds to $d = D^{\mathrm{obs}}$. See Appendix~\ref{app:proof} for a detailed discussion.

\begin{algorithm}[ht]
  \SetKwInOut{Input}{Input}
  \SetKwInOut{Output}{Output}
  \Input{Test statistic $T = T(Y(d), d)$; observed assignment $D^{\mathrm{obs}}$; observed outcome $Y^{\mathrm{obs}}$; treatment assignment mechanism $P$; significance level $\alpha$.} 
    
  \For{$r=1$ \KwTo $R$}{
     Randomly sample $d^r \sim P$ and compute $T_r \equiv T(Y(D^{\mathrm{obs}}), d^r, D^{\mathrm{obs}})$.\\
     Compute $T^{\mathrm{obs}}_r \equiv T(Y(D^{\mathrm{obs}}), D^{\mathrm{obs}}, d^r)$. 
    }
    
  \Output{\textit{p}-value: $\hat{pval}^{\mathrm{pair}} = \frac{1 + \sum_{r=1}^R 1\{T_r \geq T^{\mathrm{obs}}_r\}}{1 + R}$.}
  
  \caption{PIRT \textit{p}-value estimation} \label{algo:pair}
\end{algorithm}

\paragraph{Running Example Continued.}

Using the difference-in-mean estimator as before,
\[
T(Y(D^{\mathrm{obs}}), D^{\mathrm{obs}}, D) = 
\bar{Y}_{\mathbb{I}(D)}(D^{\mathrm{obs}})_{\{i: D^{\mathrm{obs}} \in \mathcal{D}_i(0)/\mathcal{D}_i(1)\}} 
- 
\bar{Y}_{\mathbb{I}(D)}(D^{\mathrm{obs}})_{\{i: D^{\mathrm{obs}} \in \mathcal{D}_i(1)\}}.
\]

\begin{table}[ht]
\centering
\caption{PIRT in the Example}
\label{tab:PIRT}
\setlength\tabcolsep{6pt}
\renewcommand{\arraystretch}{0.935}
{\small
\begin{tabular}{c|>{\columncolor{red}}c|>{\columncolor{blue}}c|>{\columncolor{brown}}c|>{\columncolor{brown}}c|c|c|c}
\hline \hline
Assignment $D$ & \multicolumn{4}{c|}{Potential Outcome $Y_i$} & \multicolumn{2}{c|}{\scriptsize $\bar{Y}_{\mathbb{I}(D)}(D^{\mathrm{obs}})$} & {\scriptsize $T(Y(D^{\mathrm{obs}}), D^{\mathrm{obs}}, D)$} \\ \hline
& $i_1$ & $i_2$ & $i_3$ & $i_4$ & {\scriptsize ${\{i: D^{\mathrm{obs}} \in \mathcal{D}_i(0)/\mathcal{D}_i(1)\}}$} & {\scriptsize ${\{i: D^{\mathrm{obs}} \in \mathcal{D}_i(1)\}}$} &  \\
\hline \rowcolor{gray}
$(1,0,0,0)$ &  2  &  4  &  3 &   2 & 4 & 2.5 &  1.5 \\ 
\hline
$(0,1,0,0)$  &  \color{red}{?}  & \color{red}{?}  &   $3$ &  2 & \color{red}{?} & 2.5 & \color{red}{2}\\
\hline 
$(0,0,1,0)$  &   \color{red}{?} &  4 & \color{red}{?} &  2 & 4 & 2 & 2 \\ 
\hline
$(0,0,0,1)$ &  \color{red}{?} &   4 &  3 & \color{red}{?} & 4 & 3 & 1 \\
\hline \hline
\end{tabular}
}\\
\notes{
\textbf{Notes:} Assignment $D$ includes all potential assignments, with the first row representing the observed assignment $D^{\mathrm{obs}}$. 
Potential Outcome $Y_i$ is the potential outcome of each unit under the null $H^0_0$, with red question marks indicating missing values. 
Unit $i_1$ does not belong to either the neighborhood set or the control set under $D^{\mathrm{obs}}$, so the column is marked red. 
Unit $i_2$ is in the distance interval $(0,1]$ under $D^{\mathrm{obs}}$, so the column is marked blue. 
Units $i_3$ and $i_4$ are in the distance interval $(1,\infty)$ under $D^{\mathrm{obs}}$, so those columns are marked brown. 
$T(Y(D^{\mathrm{obs}}), D^{\mathrm{obs}}, D)$ is calculated as the mean of non-missing potential outcomes in the blue columns minus the mean of non-missing potential outcomes in the brown columns.
}
\end{table}

\begin{figure}[ht]
    \centering
    \caption{Illustration of Imputable Neighbor and Control Units for $T(Y(D^{obs}), D^{obs}, D)$}
    \label{fig:PIRT}
    % Panel (a)
    \begin{subfigure}{0.22\textwidth}
        \centering
        \begin{tikzpicture}[node distance=1cm]
            \node[rectangle, draw=green!60, color=red] (i1) at (3.8,-0.8) {$i_1$};
            \node[circle, draw=green!60, color=red] (i1) at (3.8,-0.8) {$i_1$};
            \node[color=blue] (i2) at (5.2,-0.8) {$i_2$};
            \node[color=brown] (i3) at (5.2,-2.2) {$i_3$};
            \node[color=brown] (i4) at (3.8,-2.2) {$i_4$};
            % Edges
            \draw[black, ultra thick] (4,-1) -- (5,-1) (5,-2) -- (4,-2);
        \end{tikzpicture}
        \caption{$D^{obs}$: $i_1$, $D$: $i_1$}
    \end{subfigure}
    \hfill
    % Panel (b)
    \begin{subfigure}{0.22\textwidth}
        \centering
        \begin{tikzpicture}[node distance=1cm]
            \node[circle, draw=green!60, color=red] (i1) at (3.8,-0.8) {$i_1$};
            \node[rectangle, draw=green!60, color=red] (i2) at (5.2,-0.8) {$i_2$};
            \node[color=brown] (i3) at (5.2,-2.2) {$i_3$};
            \node[color=brown] (i4) at (3.8,-2.2) {$i_4$};
            % Edges
            \draw[black, ultra thick] (4,-1) -- (5,-1) (5,-2) -- (4,-2);
        \end{tikzpicture}
        \caption{$D^{obs}$: $i_1$, $D$: $i_2$}
    \end{subfigure}
    \hfill
    % Panel (c)
    \begin{subfigure}{0.22\textwidth}
        \centering
        \begin{tikzpicture}[node distance=1cm]
            \node[circle, draw=green!60, color=red] (i1) at (3.8,-0.8) {$i_1$};
            \node[color=blue] (i2) at (5.2,-0.8) {$i_2$};
            \node[rectangle, draw=green!60, color=red] (i3) at (5.2,-2.2) {$i_3$};
            \node[color=brown] (i4) at (3.8,-2.2) {$i_4$};
            % Edges
            \draw[black, ultra thick] (4,-1) -- (5,-1) (5,-2) -- (4,-2);
        \end{tikzpicture}
        \caption{$D^{obs}$: $i_1$, $D$: $i_3$}
    \end{subfigure}
    \hfill
    % Panel (d)
    \begin{subfigure}{0.22\textwidth}
        \centering
        \begin{tikzpicture}[node distance=1cm]
            \node[circle, draw=green!60, color=red] (i1) at (3.8,-0.8) {$i_1$};
            \node[color=blue] (i2) at (5.2,-0.8) {$i_2$};
            \node[color=brown] (i3) at (5.2,-2.2) {$i_3$};
            \node[rectangle, draw=green!60, color=red] (i4) at (3.8,-2.2) {$i_4$};
            % Edges
            \draw[black, ultra thick] (4,-1) -- (5,-1) (5,-2) -- (4,-2);
        \end{tikzpicture}
        \caption{$D^{obs}$: $i_1$, $D$: $i_4$}
    \end{subfigure}
    \notes{\textbf{Notes:} Treated units in $D^{obs}$ are marked with red circles and determine the neighbor units in the interval $(0,1]$ and control units in $(1,\infty)$. Treated units in $D$ are marked with red rectangles and determine the imputable units.}
\end{figure}
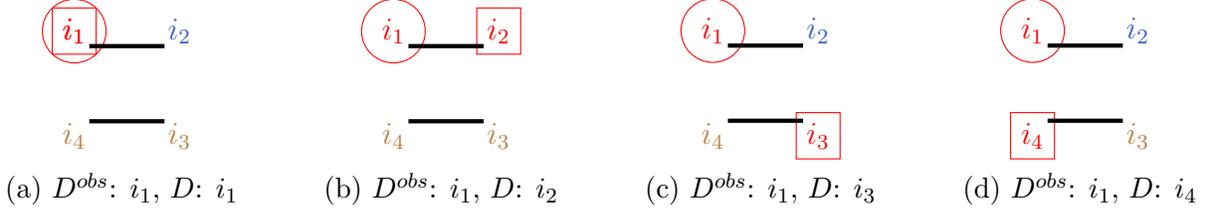

As shown in Figure~\ref{fig:PIRT} and Table~\ref{tab:PIRT}, for each treatment assignment $D$, the test statistic is calculated as the mean value of $i_2$ (excluding missing values) minus the mean value of $i_3$ and $i_4$ (excluding missing values). Based on Tables~\ref{tab:PIRT} and~\ref{tab:pair_stat}, I can construct Table~\ref{tab:pair_compare}, where each row represents the values used to compare and construct the \textit{p}-value for each $(D^{\mathrm{obs}}, D)$ pair.

\begin{sloppypar}
Only when $D$ involves treating units $i_1$ or $i_2$ does $T(Y(D^{obs}), D, D^{obs}) \geq T(Y(D^{obs}), D^{obs}, D)$. Hence, $pval^{pair} = 2/4$. In practice, similar to \citet{guan2023conformal}, using $1/2$ to discount the number of equalities can reduce the \textit{p}-value without compromising test validity. Additionally, in simulation experiments, using a uniform random number multiplied by the number of equalities also maintains test validity.
\end{sloppypar}

\begin{table}[ht]
\centering \caption{Pairwise Comparison for PIRT} \label{tab:pair_compare}
  \setlength\tabcolsep{10pt}
    \renewcommand{\arraystretch}{0.935}
    {\small \begin{tabular}{c|c|c|c|c|c|c}
    \hline \hline
    Assignment $D$ & \multicolumn{4}{c|}{Potential Outcome $Y_i$} &  {\scriptsize $T(Y(D^{obs}), D, D^{obs})$} & {\scriptsize $T(Y(D^{obs}), D^{obs}, D)$}  \\ \hline
    & $i_1$ & $i_2$ & $i_3$ & $i_4$  &  & \\
     \hline \rowcolor{gray}
 $(1,0,0,0)$ &  2  &  4  & 3 & 2 & 1.5  & 1.5 \\ 
 \hline
  $(0,1,0,0)$  & \color{red}{?}  & \color{red}{?}  & $3$ & 2 & \color{red}{2} & \color{red}{2} \\
    \hline 
 $(0,0,1,0)$  &  \color{red}{?} &  4 & \color{red}{?} & 2 & -2 & 2 \\ 
 \hline
   $(0,0,0,1)$ &   \color{red}{?} &  4 & 3 & \color{red}{?} & -1 & 1\\
  \hline \hline
  \end{tabular} }\\
  \notes{ \textbf{Notes:} Assignment $D$ includes all potential assignments, with the first row representing the observed assignment $D^{obs}$. Potential Outcome $Y_i$ is the potential outcome of each unit under the null $H^0_0$, with red question marks indicating missing values. $T(Y(D^{obs}), D, D^{obs})$ are test statistics under different $D$ while fixing $D^{obs}$ for imputable units, with the same values as in Table \ref{tab:naive}. $T(Y(D^{obs}), D^{obs}, D)$ are test statistics under different $D$ for imputable units, with the same values as in Table \ref{tab:PIRT}.}
\end{table}

Similar to \citet{guan2023conformal}, for non-directional tests, the absolute value of the test statistic can be used. For directional tests, the statistic can be applied to test for positive effects, while the negation of the statistic can be used to test for negative effects.

\paragraph{Comparison to the CRTs}

When testing under a partially sharp null hypothesis, the \textit{p}-values constructed in Definition~\ref{def:pvalue_pair} align closely with those from CRTs if we interpret $\mathbb{I}(D^{\mathrm{obs}})$ as a focal unit set and $\{0,1\}^N$ as a focal assignment set. The pair $(\mathbb{I}(D^{\mathrm{obs}}), \{0,1\}^N)$ represents a broader conditioning event than the traditional conditioning sets used in CRTs. Depending on whether the additional potential units and assignments contribute meaningful information, this broader conditioning may or may not yield higher statistical power.

In settings where a conditioning event can be specified over all imputable units in $\mathbb{I}(D^{\mathrm{obs}})$, as demonstrated by \citet{basse2024}, CRTs with a well-defined focal assignment set may allow for more targeted comparisons, potentially increasing power. Nevertheless, in scenarios where constructing a suitable conditioning event is infeasible or would result in only a limited number of focal units and assignments, the PIRT framework may provide a more practical alternative. More generally, when including all assignments from $\{0,1\}^N$ is suboptimal, combining elements of PIRTs and CRTs may further improve power by focusing on more relevant test statistics and selected assignments \citep{lehmann2006, Hennessy2015ACR}. Exploring this integration presents a promising direction for future research aimed at optimizing power through the flexibility of both PIRTs and CRTs.

\paragraph{Selection of $\epsilon_s$ and $\epsilon_c$}
\begin{sloppypar}
In practice, researchers may wish to test multiple distance levels $\epsilon_s$ rather than a single one, in order to estimate the boundary of interference. A sequential testing approach—beginning with the smallest $\epsilon_s$ and increasing until the test fails to reject—can be used to estimate this boundary while automatically controlling the family-wise error rate, without the need for further adjustment of the significance level. This procedure is especially valuable when spillover effects are positive, as it enables policymakers to design more cost-effective interventions. Conversely, if spillover effects are negative, identifying their range can inform assessments of overall policy effectiveness. See Appendix~\ref{app:multi} for a detailed discussion of this framework and considerations for sequential testing.
\end{sloppypar}

\section{Empirical Application: \citet{blattman2021}} \label{sec:apply}

In 2016, a large-scale experiment was conducted in Bogot\'a, Colombia, as described by \citet{blattman2021}. The study covered 136{,}984 street segments, of which 1{,}919 were identified as crime hotspots. Among these hotspots, 756 were randomly assigned to a treatment involving increased daily police patrolling—from 92 to 169 minutes per day over eight months. The experiment also included a secondary intervention aimed at enhancing municipal services, though this is peripheral to the primary focus of my empirical application. The key outcome of interest is the number of crimes per street segment, including both property crimes and violent crimes (such as assault, rape, and murder).

\begin{figure}[ht]
    \centering
    \caption{Map of Experimental Sample and Treatment Conditions}
    \label{fig:hotspots}
    % Left Panel (a)
    \begin{subfigure}{0.45\linewidth}
        \centering
        \includegraphics[width=\linewidth]{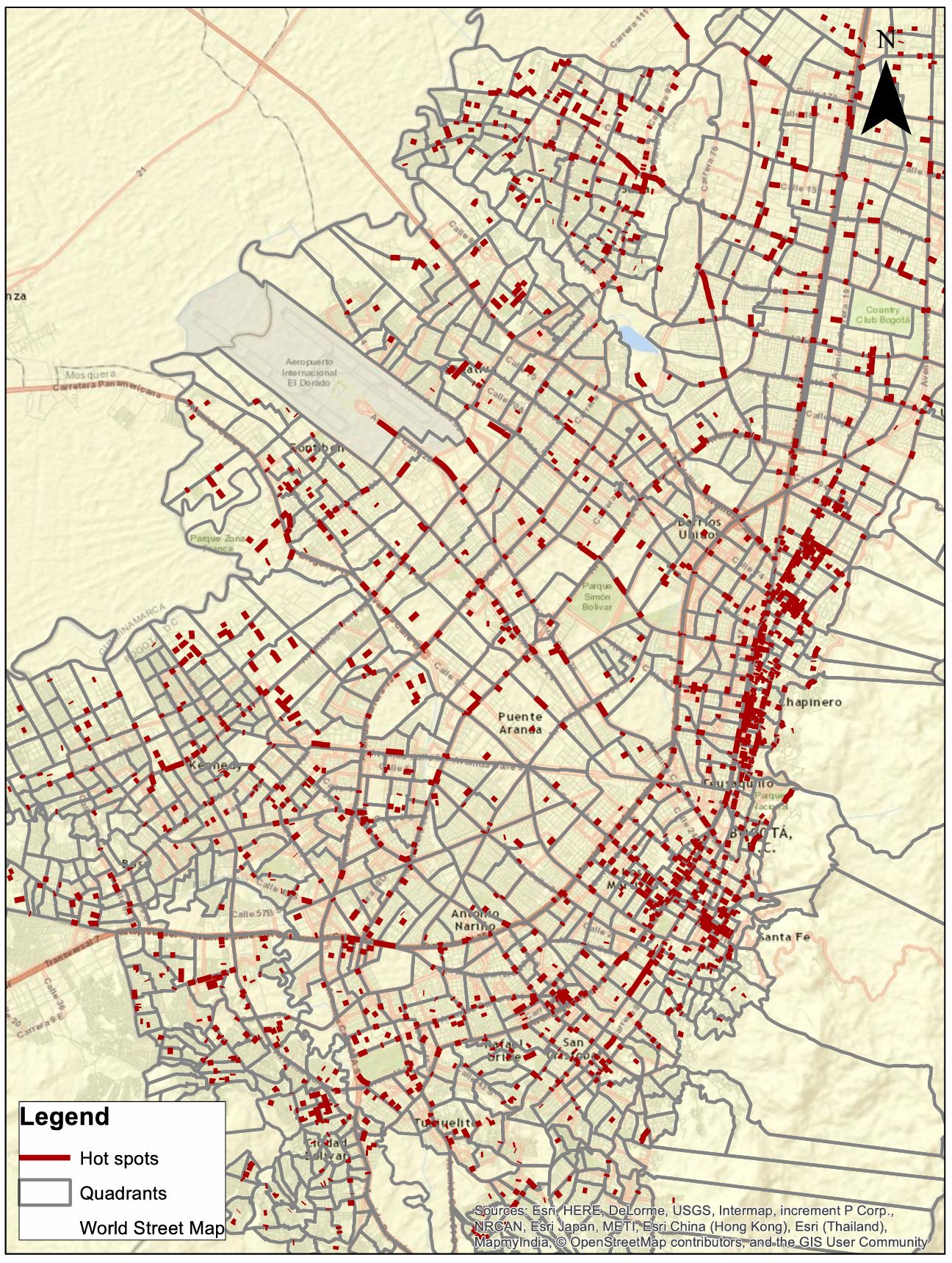}
        \caption{Experimental Sample Map}
        \label{fig:hotspots_a}
    \end{subfigure}
    \hfill
    % Right Panel (b)
    \begin{subfigure}{0.45\linewidth}
        \centering
        \includegraphics[width=\linewidth]{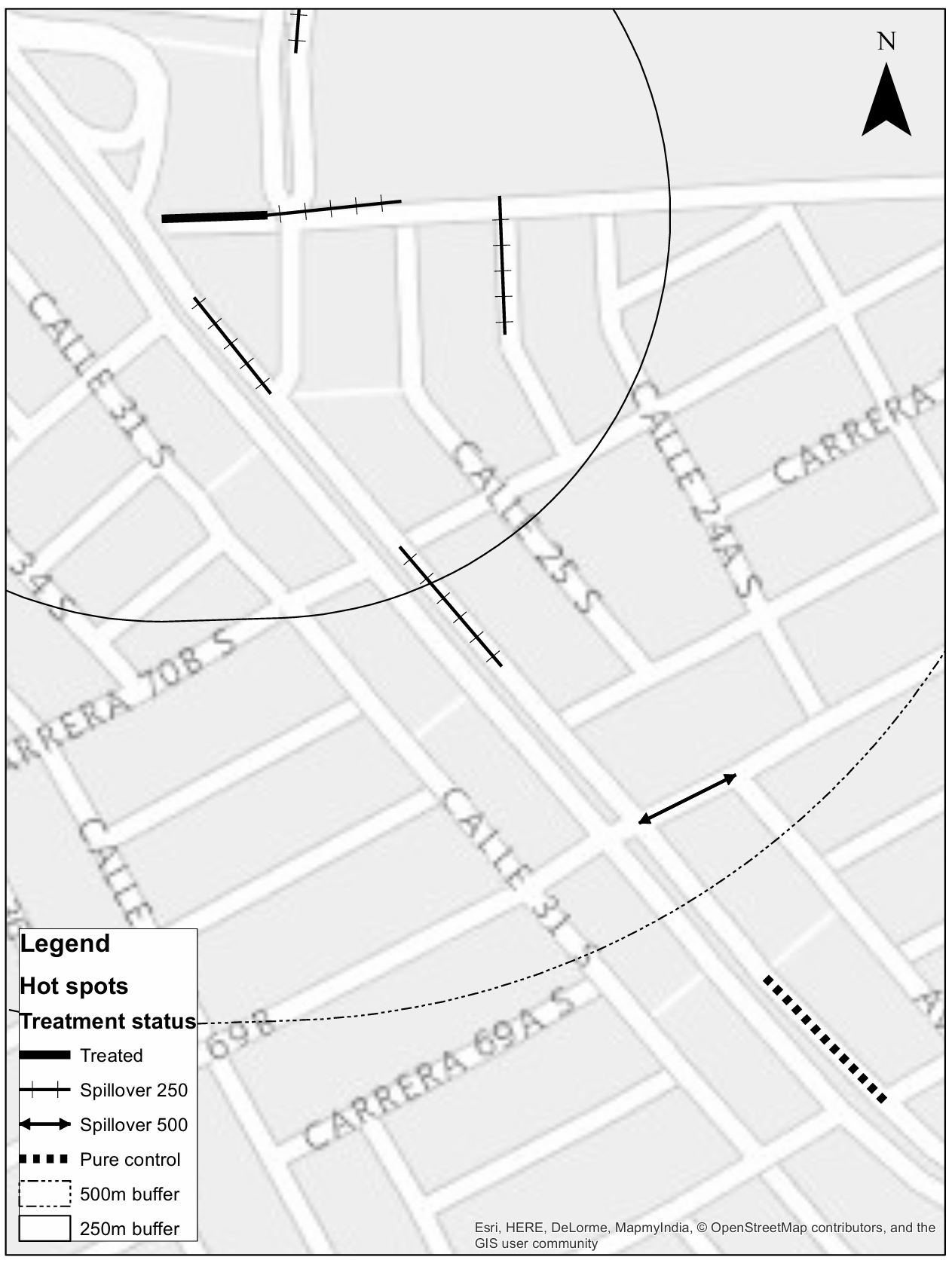}
        \caption{Assignment to Experimental Conditions}
        \label{fig:hotspots_b}
    \end{subfigure}
    \vspace{0.5em}
    \begin{minipage}{0.95\linewidth}
        \footnotesize
        \textbf{Notes:} Panel (a) displays a map of the experimental sample, with hotspot street segments marked in red. Panel (b) shows an example of the assignment to the four experimental conditions. Source: \citet{blattman2021}.
    \end{minipage}
\end{figure}

Figure~\ref{fig:hotspots_a} shows the distribution of hotspots, with many located in close proximity to one another. While only 756 street segments received the treatment, every segment potentially experienced spillover effects, resulting in a dense network of possible interactions. This complexity complicates the use of cluster-robust standard errors for addressing unit correlation.

In evaluating the total welfare impact of the policy, it is essential to determine whether interference occurred following treatment assignment—such as crime displacement or deterrence in neighboring areas. To this end, I address three key considerations: (1) whether interference exists; (2) if present, whether it manifests as displacement or deterrence; and (3) the distance at which this interference is effective. Given the complexity of modeling correlations among units in such a dense network, testing a partially sharp null hypothesis, as proposed by \citet{blattman2021} and \citet{Puelz2022}, is particularly relevant.
\begin{sloppypar}
For this analysis, I specify a sequence of distance thresholds $(\epsilon_0, \epsilon_1, \epsilon_2, \epsilon_3) = (0, 125, 250, 500)$ for $K=3$, where the interval $(500, \infty)$ serves as a pure control group—containing units with no treated neighbors within 500 meters. Figure~\ref{fig:hotspots_b} illustrates an example of these distance intervals, as identified in \citet{blattman2021}.
\end{sloppypar}

Table~\ref{tab:descriptive} presents descriptive statistics for the number of crimes observed during the intervention period. The t-statistics from t-tests comparing each pair of columns reveal two salient findings: first, treated hotspots experienced significantly fewer crimes; and second, non-hotspot areas reported progressively fewer crimes the farther they were from any treated unit. However, the exceptionally high values of the t-statistics warrant caution in interpreting these results as evidence of a displacement effect. As discussed previously, standard errors may be underestimated, and units at different distances from treated areas may not be directly comparable. Both factors could contribute to the elevated t-statistics observed in the table.

\begin{table}[ht]
\centering \caption{Descriptive Statistics During the Intervention} \label{tab:descriptive}
\scalebox{0.9}{ 
    \setlength\tabcolsep{8pt} 
    \renewcommand{\arraystretch}{0.85} 
    {\small
    \begin{tabular}{lcccccc}
    \hline \hline
    \textbf{Stats} & \multicolumn{2}{c}{\textbf{Crime hotspots}} & \multicolumn{4}{c}{\textbf{Non-hotspots (distance to treated units)}} \\ \hline
     & Treated & Non-treated & $(0m, 125m]$ & $(125m, 250m]$  & $(250m, 500m]$  & $(500m, \infty)$ \\
     \hline
   Obs.  &  756 &  1,163 & 24,571 & 32,034 & 45,147 & 33,313 \\
   [1em]  
\multicolumn{7}{l}{\textbf{\# of total crimes}} \\
  Mean  &  0.935  & 1.311 & 0.378 & 0.294  & 0.242 & 0.180 \\ 
  SD  &  1.519  & 2.332 & 1.006 & 0.921  & 0.736 & 0.602 \\ 
  Max  &  12  & 43 & 33 & 40  & 25 &  31  \\ 
  [1em]
  \% of $>0$  &  44.84  & 53.22 & 23.17 & 19.01  & 16.69 &  13.13  \\ 
  [1em]
t-stat of t-test & \multicolumn{2}{c}{-3.93} & \multicolumn{4}{c}{10.34  \hspace{1.5cm} 8.60 \quad\hspace{1.5cm} 12.62} \\
 [1em]
\multicolumn{7}{l}{\textbf{\# of property crimes}} \\
 Mean  &  0.712  & 1.035 & 0.262 & 0.195  & 0.158 & 0.111 \\ 
 SD  &  1.269  & 2.099 & 0.778 & 0.683  & 0.555 & 0.441 \\ 
 Max  &  12  & 40 & 32 & 36  & 21 &  27 \\ 
 [1em]
\% of $>0$  &  38.36  & 45.66 & 17.44 & 13.96  & 11.90 &  8.86  \\
 [1em]
t-stat of t-test & \multicolumn{2}{c}{-3.81} & \multicolumn{4}{c}{10.93  \hspace{1.5cm} 8.26 \quad\hspace{1.5cm} 12.83} \\
    [1em]
\multicolumn{7}{l}{\textbf{\# of violent crimes}} \\
  Mean  &  0.224  & 0.276 & 0.115 & 0.099  & 0.084 & 0.069 \\ 
  SD  &  0.593  & 0.650 & 0.467 & 0.473  & 0.376 & 0.334\\ 
  Max  &  5  & 6 & 17 & 40  & 13 & 11\\ 
  [1em]
\% of $>0$  &  16.40  & 20.29 & 8.59 & 7.29  & 6.51 &  5.47  \\
 [1em]
t-stat of t-test & \multicolumn{2}{c}{-1.79} & \multicolumn{4}{c}{4.23  \hspace{1.5cm} 4.75 \quad\hspace{1.5cm} 5.79} \\
  \hline \hline
  \end{tabular} } 
  }\\ 
\notes{\textbf{Notes:} This table presents descriptive statistics for crime data during the intervention, divided into two categories: crime hotspots (treated and non-treated) and non-hotspot areas, which are further grouped by their distance from the treated units. The statistics cover three types of crimes: total crimes, property crimes, and violent crimes. For each group, the table provides the mean, standard deviation (SD), maximum (Max), and the percentage of units with positive crimes (\% of $>0$). The t-statistic values (t-stat of t-test) represent the results from t-tests comparing the difference in means between treated versus non-treated units within crime hotspots, non-hotspot units within $(0m, 125m]$ versus $(125m, 250m]$, non-hotspot units within $(125m, 250m]$ versus $(250m, 500m]$, and non-hotspot units within $(250m, 500m]$ versus $(500m, \infty)$.}
\end{table}

The original study estimated a negative treatment effect and conducted inference using Fisher Randomization Tests (FRTs) under a sharp null hypothesis of no effect. \citet{blattman2021} reported no significant displacement effect for violent crimes and a marginally significant displacement effect for property crimes.\footnote{The treatment effect for violent crime was significant, but property crime effects were insignificant.} However, as previously discussed, p-values from t-tests may not adequately capture the extent of interference, and employing FRTs to test partially sharp null hypotheses may not be valid in this context. Consequently, it is important to consider how these conclusions might differ if a valid testing approach is employed.

%As a first step, \citet{blattman2021} employed an approach similar to the inner vs. outer ring strategy to identify the pure control group for analysis. However, as discussed in subsection \ref{sec:null} and in \citet{pollmann2023causal}, this method does not guarantee size control. Therefore, this section revisits the analysis to identify the pure control group and determine the neighborhood of interference using methods introduced in Section \ref{sec:multi}. \textbf{Might need to rewrite a bit?}

\subsection{Power Comparison of Spatial Interference: A Simulation Study} \label{sec:simu}

To compare the power and validity of competing inference methods under spatial interference, I conduct a simulation study designed to preselect the most promising approach for the main analysis. The simulation sample consists of 1,000 units, comprising 20 hotspots and 7 randomly treated units, closely matching the proportions observed in the original Bogot\'a study. I focus on two distance thresholds, using $(\epsilon_0, \epsilon_1, \epsilon_2) = (0, 0.1, 0.2)$.

To approximate the Bogot\'a context, I calibrate the schedule of potential outcomes using gamma distributions that match the observed mean and variance of total crimes. A negative treatment effect of $-1$ is imposed, ensuring non-negative crime counts for treated units. Additionally, a \emph{decreasing displacement effect} is introduced via a positive parameter $\tau$: this models spatial spillovers such that the indirect effect of treatment decays with distance from the treated unit, reflecting the empirical pattern observed in the data and representing the primary focus of the analysis.

The partially sharp null hypothesis for $k = 0$ and $1$ is given by
\[
H^{\epsilon_k}_0: Y_i(d) = Y_i(d^\prime) \quad \text{for all } i \in \{1, \dots, N\}, \text{ and any } d, d^\prime \in \mathcal{D}_i(\epsilon_k).
\]

In the analysis, I compare four methods: (1) the classic FRT, applied under the sharp null hypothesis of no effect, as in \citet{blattman2021} for inference on spillover effects; (2) the biclique Conditional Randomization Test (CRT) proposed by \citet{Puelz2022}, which serves as a benchmark due to its strong power in simulations under general interference; (3) the PIRT with rejection at the $\alpha/2$ level, ensuring validity under the worst-case scenario; and (4) the PIRT with rejection at the nominal $\alpha$ level.

Two main criteria guide the choice of test. First, under no spillover effect ($\tau = 0$), the partially sharp null hypothesis should be rejected no more than 5\% of the time, maintaining type I error control. Second, when a spillover effect is present ($\tau > 0$), the test should maximize rejection of the null, i.e., maximize power. To assess power, I consider 50 equally spaced values of $\tau$ between 0 and 1, conducting 2,000 simulations for each value, and compute the average rejection rate for each method. Due to the computational challenge posed by the biclique CRT, I first draw $5{,}000$ assignments to approximate the full space of potential treatment assignments, then construct the conditioning event.\footnote{The current code still requires over four hours to construct the conditioning events for the simulations.} See Appendix \ref{app:algo} of the Supplemental Material for details.

\begin{figure}[!htbp]
    \centering
    \caption{Power Comparison of Testing Methods for Different Hypotheses}
    \label{fig:power_simu}
    \includegraphics[width=1\textwidth]{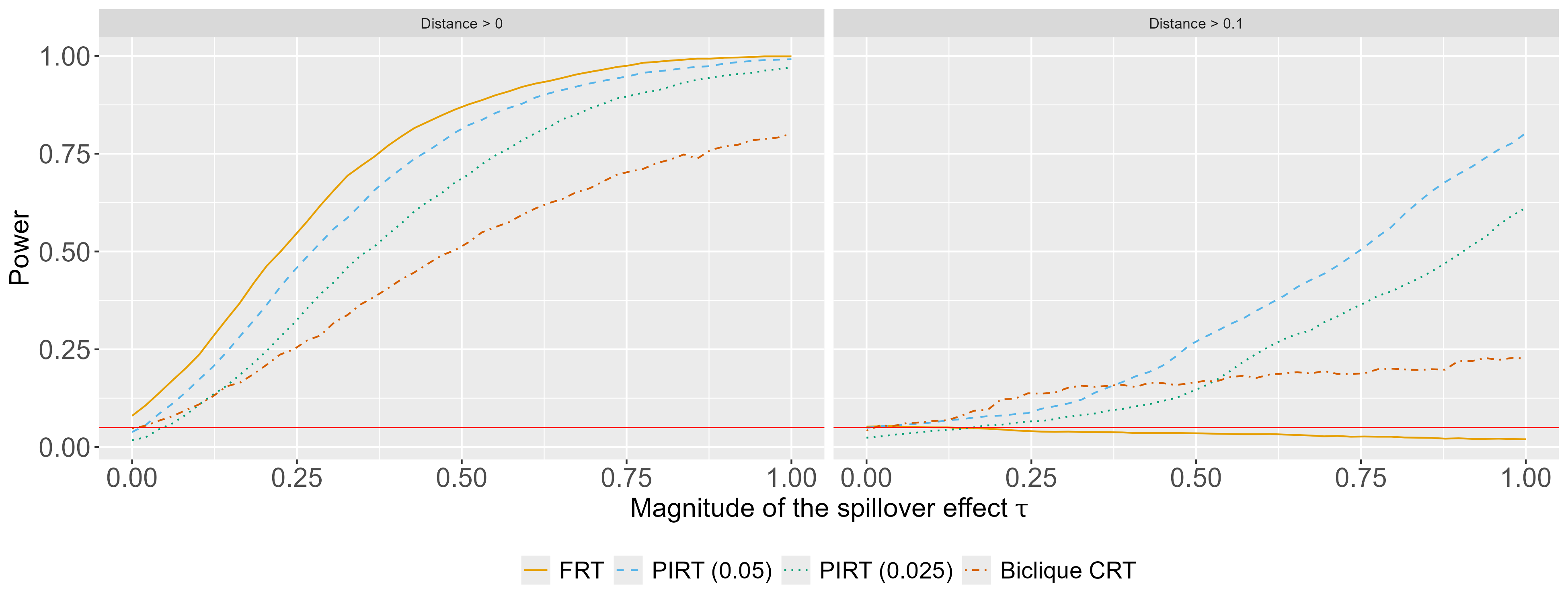}
    \notes{\textbf{Notes:} The left panel shows power for testing $H^0_0$, while the right panel shows power for $H^{0.1}_0$. The red line marks the size $\alpha = 0.05$. “PIRT (0.025)” denotes PIRT with rejection at $\alpha/2$; “PIRT (0.05)” at $\alpha$.}
\end{figure}

Figure \ref{fig:power_simu} (left panel) shows that the FRT over-rejects the true partially sharp null when $\tau = 0$, consistent with \citet{athey2018}'s observation that testing the sharp null of no effect is invalid for partially sharp nulls. In my simulation, with only seven treated units (0.7\% of the total), the FRT rejection rate is around 10\%. Notably, the unadjusted PIRT at level $\alpha$ maintains good size control, indicating that the $\alpha/2$ threshold is mainly a worst-case guarantee and can be conservative, with rejection rates below 5\%. The biclique CRT is also valid, with a rejection rate near 5\%.

Regarding power, I exclude the FRT from comparison due to its invalidity. The unadjusted PIRT ($\alpha$) demonstrates the best performance, outperforming other methods across all effect sizes $\tau$. Among methods with theoretical size control, the PIRT with $\alpha/2$ rejection is optimal, though it slightly lags the biclique CRT at small $\tau$. Despite its validity, the biclique CRT's power increases slowly as the spillover effect grows: the rejection rate remains below 90\% even at $\tau=1$.

The right panel of Figure \ref{fig:power_simu} shows a contrasting pattern. First, all methods—including the FRT—maintain validity under the null. This may reflect that hotspots rarely fall into exposure levels $(0.1, 0.2]$ or $(0.2, \infty)$, so despite a negative treatment effect, its impact on the test statistic is minimal. As with $H^0_0$, both PIRT and biclique CRT exhibit rejection rates near 5\%, while the PIRT at $\alpha/2$ remains conservative.

Second, all methods show considerably lower power than for $H^0_0$. This is mainly because only 60\% of units are relevant to the partially sharp null, and the spillover effect is halved ($0.5\tau$). Nonetheless, the PIRT method still achieves appreciable power when $\tau$ is large, outperforming alternatives, especially with the unadjusted $\alpha$ rejection level. Interestingly, the FRT now under-rejects, with almost no power for any $\tau$. This occurs because the FRT's $p$-value remains large unless the observed test statistic exceeds most under randomization. However, units in $(0, 0.1]$ under the observed assignment contribute to test statistics for other assignments, and these units, subject to spillover $\tau$, cause observed statistics for $(0.1, 0.2]$ and $(0.2, \infty)$ to remain low even at high $\tau$. As a result, the FRT $p$-value remains large, explaining the lack of power.

Taken together, these results illustrate how the FRT, when applied to partially sharp nulls, may yield either over- or under-rejection, depending on the scenario. Although the biclique CRT displays power, its increase is slower than with PIRT, likely due to the challenge of finding optimal conditioning events under spatial interference. The biclique method's performance may improve with advanced computing resources, as it depends on parameter choices in biclique finding. The main advantage of my method is computational simplicity; further research could expand power comparisons among approaches.

Overall, the results favor PIRT methods, especially the unadjusted PIRT. Accordingly, I apply PIRTs to replicate the results of \citet{blattman2021}, using the non-absolute difference-in-means estimator.

\subsection{Implementation of PIRTs for Testing the Existence of a Displacement Effect}

Consider the experimental setting described in \citet{blattman2021}, where the observed treatment assignment is denoted by $D^{\mathrm{obs}}$. Following their approach, let $Y$ denote the number of crimes, and let $S(D^{\mathrm{obs}})$ be an indicator for whether a unit is within 125 meters of any treated unit under the observed assignment. This proximity indicator is determined by $D^{\mathrm{obs}}$ and would change under a different assignment $D$. Although additional covariates may be included in the regression, the primary test statistic is the coefficient from regressing $Y$ on $S(D)$.

To implement the PIRT for testing the partially sharp null hypothesis in practice, proceed as follows:

\begin{enumerate}
    \item[\textbf{1.}] Randomly reassign treatments $D$ to the units.
    \item[\textbf{2.}] For each reassignment $D$, identify the subsample of units untreated under both $D^{\mathrm{obs}}$ and $D$. These are the \emph{imputable units}, since their potential outcomes are unaffected by treatment under either assignment.
    \item[\textbf{3.}] Compute the coefficient $\beta$ from regressing $Y$ on $S(D^{\mathrm{obs}})$ within the subsample of imputable units.
    \item[\textbf{4.}] Compute the coefficient $\beta'$ from regressing $Y$ on $S(D)$ within the same subsample of imputable units.
\end{enumerate}

Steps 3 and 4 construct the pairwise imputable statistics, relying exclusively on units untreated under both the observed and randomized assignments.

For hypothesis testing regarding displacement effects, the PIRT computes the $p$-value as the fraction of reassignments $D$ for which $\beta' \geq \beta$. The null hypothesis of no displacement effect is rejected if this $p$-value is less than or equal to $\alpha/2$. Simulation results indicate that using $\alpha$ as the rejection threshold is also empirically valid. The method can be readily adapted to two-sided tests by comparing $|\beta'| \geq |\beta|$, or to tests for deterrence effects by comparing $-\beta' \geq -\beta$.

\subsection{PIRT on Actual Data} \label{sec:rep}

I apply the proposed method to the publicly available dataset from \citet{blattman2021}, which provides street-level treatment assignments and distance intervals at thresholds of 125, 250, and 500 meters. The dataset also includes $1,000$ pseudo-randomized treatment assignments and their associated distance intervals, as used in the original study for randomization inference. However, the absence of precise longitude and latitude data for street segments precludes extending randomization testing beyond these $1,000$ assignments.

Given the high frequency of zero outcomes, I consider both an indicator for any crime occurrence and the raw number of crimes as outcome variables. Table~\ref{tab:displace_prop} reports results for these outcomes across different distance thresholds.

\begin{table}[ht]
  \centering 
  \caption{\textit{p}-Values: Testing the Displacement Effect at Different Distances}  
  \label{tab:displace_prop}
  \setlength\tabcolsep{9pt}
  \renewcommand{\arraystretch}{1.2}
{\small 
\begin{tabular}{l|ccc}
  \hline \hline
   & \multicolumn{3}{c}{Unadjusted \textit{p}-values}     \\
      & $(0m, \infty)$ & $(125m, \infty)$ & $(250m, \infty)$   \\ 
  \hline
\textit{Violent crime} &  &  &     \\
Indicator of $>0$ & \textit{0.027} & 0.812 & 0.060   \\ 
\# of crimes & \textit{0.047} & 0.546 & \textit{0.045}   \\ 
   \hline
\textit{Property crime} &  &  &     \\
 Indicator of $>0$ & 0.390 & 0.466 & 0.486    \\ 
\# of crimes & 0.325 & 0.346 & 0.394    \\ 
   \hline \hline
\end{tabular}  
} 
\notes{ \textbf{Notes:} The table shows the impact of intensive policing on violent and property crime. ``Indicator of $>0$'' refers to an indicator for any crime occurrence, while ``\# of crimes'' represents the raw number of reported crimes. \textit{p}-values are constructed using the difference-in-means estimator as the test statistic.} 
\end{table}

In the main analysis, I implement the PIRT using the difference-in-means estimator as the test statistic. A key advantage of this framework is its validity regardless of the chosen statistic. Nonetheless, researchers may incorporate covariates to improve power or examine heterogeneity; see Appendix~\ref{app:covar_results} for further discussion and empirical results with covariate adjustment.

Table~\ref{tab:displace_prop} reveals evidence of significant displacement effects for violent crime, but not for property crime. This finding stands in contrast to the original study, which found no significant displacement for violent crime. After correcting for multiple hypothesis testing using Algorithm~\ref{algo:multi}, the PIRT detects a significant short-range displacement effect (within 125 meters) at the 10\% level when using the difference-in-means estimator, and potentially at the 5\% level if rejecting directly at level $\alpha$, as supported by simulation evidence. For property crime, no spillover effects are detected at any distance.

Further, for violent crime, there is evidence of additional spillover effects beyond 250 meters, with an unadjusted \textit{p}-value of $0.045$ for the $(250\text{m}, \infty)$ interval when using the number of crimes as the outcome in PIRT. This suggests the existence of two types of offenders: high-risk offenders who relocate farther away from the intervention site, and lower-risk offenders who are displaced to nearby areas.

\begin{sloppypar}
To test this hypothesis, I further disaggregate violent crimes into socially costly crimes--homicides and sexual assaults--and other violent crimes. Table \ref{tab:violent_prop} presents results that investigate heterogeneous displacement patterns within violent crimes, revealing distinct interference effects. Specifically, socially costly crimes contribute to spillover effects beyond 250 meters but not within 125 meters. High-risk offenders involved in homicides and sexual assaults tend to relocate farther away in response to intensive policing, rather than being displaced to nearby neighborhoods. This finding underscores the dynamic responses of different types of offenders to hot-spot policing. 
\end{sloppypar}

To the best of my knowledge, this is the first causal evidence of a displacement effect extending to more distant areas rather than proximate neighborhoods.\footnote{This finding suggests that the distance interval $(500m, \infty)$ may serve as a more appropriate control group than the $(250m, \infty)$ interval used by \citet{blattman2021}.} However, after adjusting for multiple hypothesis testing using Algorithm \ref{algo:multi}, these results are no longer statistically significant. Applied researchers should interpret these findings with caution in future studies. 

\begin{table}[ht]
  \centering 
  \caption{\textit{p}-Values: Heterogeneous Patterns Within Violent Crimes}  
  \label{tab:violent_prop}
  \setlength\tabcolsep{9pt}
  \renewcommand{\arraystretch}{1.2}
  {\small 
  \begin{tabular}{lccc}
  \hline \hline
   & \multicolumn{3}{c}{Unadjusted \textit{p}-values}     \\
   & $(0m, \infty)$ & $(125m, \infty)$ & $(250m, \infty)$   \\ 
  \hline
  \multicolumn{4}{l}{\textit{Homicides and sexual assaults}} \\
  Indicator of $>0$ & 0.274 & 0.864 & 0.065   \\ 
  \# of crimes & 0.417 & 0.815 & \textit{0.043}     \\ 
  \hline
  \multicolumn{4}{l}{\textit{Not homicides or sexual assaults}}  \\
  Indicator of $>0$ & \textit{0.030} & 0.752 & 0.097    \\ 
  \# of crimes & \textit{0.044} & 0.491 & 0.057   \\ 
  \hline \hline
  \end{tabular}  
  }
  \notes{ \textbf{Notes:} The table reports the impact of intensive policing on two types of violent crimes. Indicator of $>0$: indicator of any crime; \# of crimes: raw number of reported crimes. \textit{p}-Values constructed based on the difference-in-means estimator as the test statistic.} 
\end{table}

These results not only highlight the general applicability of the PIRT method but also provide suggestive evidence for policy implications and potential criminal motives in Bogot\'a, following the insights of \citet{blattman2021}. From a policy perspective, it remains unclear whether reallocating state resources to these hotspots has led to an overall reduction in crime. Further investigation is needed to identify the specific locations most affected by displacement so that those areas can be directly targeted.

Regarding criminal motives in Bogot\'a, a possible explanation—consistent with standard economic models of crime—is that violent crime in the city's hotspots is not purely expressive, as suggested by \citet{blattman2021}. Instead, some violent crimes, such as contract killings, may be driven by generally mobile criminal rents. By increasing the risk of detection, intensive policing deters criminals from committing crimes in specific locations, but the crimes themselves may relocate rather than be entirely prevented. In contrast, property crimes—which are often instrumental and linked to immobile criminal rents—appear to be deterred without causing further spillover effects. As \citet{blattman2021} noted, violent crimes are often considered more severe than property crimes, making displacement effects an essential consideration when evaluating the overall welfare impact of policy interventions.

\section{Conclusion} \label{sec:conclude}

This paper introduces a testing framework for detecting interference in network settings. The proposed tests offer computational simplicity over previous methods while retaining strong power and size properties, making them highly applicable for empirical research.

Theoretically, I formalize unconditional randomization testing and PIRT, addressing two primary challenges in testing partially sharp null hypotheses: only a subset of potential outcomes is imputable, and the set of units with imputable potential outcomes varies across treatment assignments. PIRT addresses the first challenge by employing pairwise imputable statistics and the second by constructing \textit{p}-values through pairwise comparisons. I prove that PIRT maintains size control, and I propose a sequential testing procedure to estimate the ``neighborhood'' of interference, ensuring control over the FWER.

Beyond network settings, PIRT may have broader applicability. For instance, \citet{Zhang2021} shows that partially sharp null hypotheses are relevant in time-staggered designs. This opens promising avenues for future research, including extending the framework to quasi-experimental settings and observational studies. In quasi-experimental designs, developing a unified framework that can be applied to time-staggered adoption, regression discontinuity, and network settings would be highly valuable \citep{Borusyak2023, morgan2021}. For observational studies, incorporating propensity score weighting to create pseudo-random treatments and conducting sensitivity analyses would be crucial, as noted by \citet{rosenbaum2020design}.

While simulations suggest that PIRT performs favorably compared to CRTs, their power properties remain unexplored. Insights from studies such as \citet{Puelz2022} on CRT power and \citet{wen2023residual} on the near-minimax optimality of minimization-based \textit{p}-values suggest that further investigation into the power of PIRT could yield valuable insights. Additionally, power may increase when PIRT is combined with CRTs in specific settings, making the construction of an optimal testing framework for interference an important direction for future research.

\newpage
\bibliographystyle{ecta}
\bibliography{refs}

%-----------------------------------------------------------
%         Appendices
%-----------------------------------------------------------
\begin{appendices}
  \renewcommand\thetable{\thesection\arabic{table}}
  \renewcommand\thefigure{\thesection\arabic{figure}}
  \renewcommand{\theequation}{\thesection.\arabic{equation}}
  \renewcommand\thelemma{\thesection\arabic{lemma}}
  \renewcommand\thetheorem{\thesection\arabic{theorem}}
  \renewcommand\thedefinition{\thesection\arabic{definition}}
    \renewcommand\theassumption{\thesection\arabic{assumption}}
  \renewcommand\theproposition{\thesection\arabic{proposition}}
    \renewcommand\theremark{\thesection\arabic{remark}}
    \renewcommand\thecorollary{\thesection\arabic{corollary}}
\setcounter{equation}{0}
\setcounter{lemma}{0}
\clearpage \baselineskip=18.0pt
\appendix
%\section{Preliminary Results for Section \ref{sec:main}} \label{apx:prelim}
% Reset numbering for definitions, theorems, and algorithms in the appendix
\counterwithin{definition}{section} % Renumber definitions by section in appendix
\counterwithin{theorem}{section}    % Renumber theorems by section in appendix
\counterwithin{algorithm}{section}  % Renumber algorithms by section in appendix
\counterwithin{table}{section}
\counterwithin{figure}{section}

\section{Proof of the Theorems} \label{app:proof}

\paragraph{Proof of Proposition \ref{prop:validity}.}
For any $d, d^\prime\in \{0,1\}^N$. Consider any $i \in \mathbb{I}(d) \cap \mathbb{I}(d^\prime)$. By Definition \ref{def:imput_units} of imputable units, under $H^{\epsilon_s}_0$, we have $Y_i(d)=Y_i(d^\prime)$. Hence, by Definition \ref{def:test_stat} of pairwise imputable statistics, $T(Y(d), d^\prime,d)=T(Y(d^\prime), d^\prime, d)$.
\qed

\paragraph{Proof of Theorem \ref{theo:validity_pair}.}

Given any $\alpha > 0$, consider the subset of assignment 
\[\mathbb{D} \equiv \{D^{obs}| pval^{pair}(D^{obs}) \leq \alpha/2\}.\] 
Therefore, we can denote $P(pval^{pair}(D^{obs}) \leq \alpha/2) = \sum_{D^{obs} \in \mathbb{D}} P(D^{obs}) = w$. Since $E_P(\phi(D^{obs})) = P(pval^{pair}(D^{obs}) \leq \alpha/2)$, to prove the theorem, we want to show $w < \alpha$.

Denote $H(D^{obs}, D)= 1\{T(Y(D^{obs}), D, D^{obs}) \geq T(Y(D), D^{obs}, D)\}$. Then, by construction, $H(D^{obs}, D)+H(D,D^{obs}) \geq 1$. 

Under $H^{\epsilon_s}_0$, by Proposition \ref{prop:validity} and Definition \ref{def:pvalue_pair} of \textit{p}-value,
\[pval^{pair}(D^{obs}) = \sum_{D \in \{0,1\}^N} H(D^{obs},D)P(D).\]

Now, consider the term \[\sum_{D^{obs} \in \mathbb{D}} \sum_{D \in \{0,1\}^N} H(D^{obs},D)P(D)P(D^{obs}).  \]

On the one hand, it equals \[\sum_{D^{obs} \in \mathbb{D}} pval^{pair}(D^{obs})P(D^{obs}) \leq (\alpha/2)(\sum_{D^{obs} \in \mathbb{D}} P(D^{obs}))=w \alpha/2. \]

On the other hand, by flipping $D$ and $D^{obs}$ in the same set $\mathbb{D}$,
\begin{align*}
    \sum_{D^{obs} \in \mathbb{D}} \sum_{D \in \mathbb{D}} H(D^{obs},D)P(D)P(D^{obs}) &= \sum_{D \in \mathbb{D}} \sum_{D^{obs} \in \mathbb{D}} H(D, D^{obs})P(D^{obs})P(D) \\
    &= \sum_{D \in \mathbb{D}} \sum_{D^{obs} \in \mathbb{D}} H(D, D^{obs})P(D)P(D^{obs}) \\
    &= \sum_{D^{obs} \in \mathbb{D}} \sum_{D \in \mathbb{D}} H(D,D^{obs})P(D)P(D^{obs}).  
\end{align*}

Hence, we would have
\begin{align*}
    \sum_{D^{obs} \in \mathbb{D}} \sum_{D \in \{0,1\}^N} H(D^{obs},D)P(D)P(D^{obs}) &\geq \sum_{D^{obs} \in \mathbb{D}} \sum_{D \in \mathbb{D}} H(D^{obs},D)P(D)P(D^{obs}) \\
    &= \sum_{D^{obs} \in \mathbb{D}} \sum_{D \in \mathbb{D}} (H(D, D^{obs}) + H(D^{obs},D))P(D)P(D^{obs})/2 \\
    &(\text{By } H(D^{obs}, D^{obs}) + H(D^{obs}, D^{obs})=2) \\
    &> \sum_{D^{obs} \in \mathbb{D}} \sum_{D \in \mathbb{D}} P(D)P(D^{obs})/2 = w^2/2. 
\end{align*}

Hence, $w^2/2 < w \alpha/2$, implying $w < \alpha$. As previously mentioned, using 1/2 to discount the number of equalities does not affect the test's validity because $H(D^{obs}, D)+H(D,D^{obs}) \geq 1$ would still hold. 

\qed

\paragraph{Too Many Potential Treatment Assignments.} When the number of units $N$ is large, there would be $2^N$ potential treatment assignments, which is a large number in practice. In such cases, given $D^{obs}$ and Algorithm \ref{algo:pair}, we can show that $\|\hat{pval}^{pair}-pval^{pair}(D^{obs})\|=O_p(R^{-1/2})$. Specifically, by $\hat{pval}^{pair} = (1+\sum^R_{r=1} 1\{T_r \geq T^{obs}_r\})/(1+R)$ and $d^r \sim P$ independently, we have $E_{d^r}\hat{pval}^{pair}= pval^{pair}(D^{obs})$ and

\[Var(\hat{pval}^{pair})= Var(1\{T_r \geq T^{obs}_r\})/(1+R) = pval^{pair}(D^{obs})(1-pval^{pair}(D^{obs}))/(1+R).\]

Hence, by Chebyshev's inequality, $\|\hat{pval}^{pair}-pval^{pair}(D^{obs})\|=O_p(R^{-1/2})$.

\section{Framework for Intersection of Null Hypotheses} \label{app:inter}

Some hypotheses of interest can be expressed as the intersection of the partially sharp null hypotheses discussed in the main text. For example, \citet{athey2018} and \citet{Puelz2022} take a more stringent approach to testing for the existence of any interference by defining the following null hypothesis regarding the extent of interference at distance $k$:

\begin{definition}[Extent of Interference for Distance \(k\) in \citet{Puelz2022}] \label{def:extent_k}
\textit{In a social network, the null hypothesis at distance \(k\) states that for all \(i = 1, \dots, N\),}
\[
Y_i(d) = Y_i(d') \quad \text{for any } d, d' \in \{0,1\}^N \text{ such that } d_j = d'_j \text{ for all } j \text{ with } d(i,j) \leq k.
\]
\end{definition}

This hypothesis asserts that a unit's outcome depends only on treatments within \(k\)-hops, not beyond. Unlike the partially sharp null in Definition~\ref{def:partial}, which requires potential outcomes to remain the same across one subset of assignments, this hypothesis allows unit \(i\)'s outcome to change whenever a nearby unit \(j\) (with \(d(i,j)\leq k\)) switches treatment status. Nevertheless, for each combination of assignment statuses within \(k\)-distance, there is a subset of assignments yielding the same outcome for \(i\). 

Moreover, the null hypothesis in Definition~\ref{def:extent_k} differs from that in the main text: each combination of treatment statuses within \(k\)-distance can be viewed as a separate partially sharp null hypothesis. The main text, by contrast, focuses on a particular partially sharp null in which all units within \(k\)-distance are untreated.

Nonetheless, the framework presented in this paper naturally extends to intersections of partially sharp null hypotheses with minor modifications. We redefine the partially sharp null hypothesis in terms of a collection \(\mathcal{D}^a = \{\mathcal{D}^a_i\}_{i=1}^N\):

\begin{definition}[Partially Sharp Null Hypothesis for \(\mathcal{D}^a\)] \label{def:partial_a}
A partially sharp null hypothesis holds if there exists a collection of subsets \(\mathcal{D}^a = \{\mathcal{D}^a_i\}_{i=1}^N\), where each \(\mathcal{D}^a_i \subsetneq \{0,1\}^N\), such that
\[
H^{\mathcal{D}^a}_0: Y_i(d) = Y_i(d') \quad \text{for all } i \in \{1, \dots, N\}, \quad \text{and any } d, d' \in \mathcal{D}^a_i.
\]
\end{definition}

\begin{definition}[Intersection of Partially Sharp Null Hypotheses] \label{def:intersect}
For each \(a \in \mathbb{F} = \{1, \dots, F\}\) and the given \(\mathcal{D}^a\), the intersection of partially sharp null hypotheses is defined as
\[
\bigcap_{a \in \mathbb{F}} H_0^{\mathcal{D}^a} := H_0^{\mathbb{F}},
\]
which is equivalent to
\[
H_0^{\mathbb{F}}: Y_i(d) = Y_i(d') \quad \text{for all } i \in \{1, \dots, N\}, \quad \text{and any } d, d' \text{ s.t. } \exists a \in \mathbb{F} \,\text{with } d, d' \in \mathcal{D}^a_i.
\]
\end{definition}

With slight adjustments to the definitions of imputable unit sets and pairwise imputable statistics, our main procedure can also test intersection null hypotheses \(H_0^{\mathbb{F}}\).

\subsection{Modifications for Testing the Intersection of Partially Sharp Null Hypotheses}

\begin{definition}[Imputable Units Set (Intersection)]\label{def:imput_units_intersect}
Given two treatment assignments \( d, d' \in \{0,1\}^N \) and an intersection of partially sharp null hypotheses \( H_0^{\mathbb{F}} \), define
\[
\mathbb{I}(d, d') 
\,\equiv\, 
\bigl\{ i \in \{1, \dots, N\} : \exists\, a \in \mathbb{F} \,\text{ s.t. }\, d, d' \in \mathcal{D}^a_i \bigr\} 
\,\subseteq\, 
\{1, \dots, N\}.
\]
as the \emph{imputable units set} under treatment assignments \( d \) and \( d' \).
\end{definition}

\begin{definition}[Pairwise Imputable Statistic (Intersection)]\label{def:test_stat_intersect}
Let 
\(
T 
\;:\; 
\mathbb{R}^N 
\times 
\{0,1\}^N 
\times 
\{0,1\}^N 
\;\longrightarrow\; 
\mathbb{R} \;\cup\;\{\infty\}
\) 
be a measurable function. We say that $T$ is \emph{pairwise imputable} if, for any $d, d' \in \{0,1\}^N$ and any pair of outcome vectors $Y, Y' \in \mathbb{R}^N$, the following holds:
\[
\text{If } Y_i = Y'_i \text{ for all } i \in \mathbb{I}(d, d'), \text{ then } T(Y, d, d') = T(Y', d, d').
\]
\end{definition}

All the propositions and theorems remain valid; the proofs are analogous to those in the main text, except that they use two treatment assignments to define the set of imputable units.

\newpage

\begin{titlingpage} 
  \emptythanks
  \title{ {Supplement to\\ \lQ {\bf Unconditional Randomization Tests for Interference}''}}
  \author{Liang Zhong\thanks{Department of Economics, Boston University, 270 Bay State Road, Boston, MA 02215 USA.\newline Email: \href{mailto:samzl@bu.edu}{samzl@bu.edu}, Website:\href{https://samzl1.github.io/}{https://samzl1.github.io/}}}
  \setcounter{footnote}{0}
  \setcounter{page}{0}

  \clearpage 
  \maketitle 
  \thispagestyle{empty} 
  \begin{center}
  This Supplemental Material consists of Appendices \ref{app:min_PIRT}, \ref{app:empty}, \ref{app:multi}, \ref{app:algo}, and \ref{app:covar} to the main text.
  \end{center}
\end{titlingpage}

\setcounter{page}{1}

\section{The Minimization-Based PIRT}\label{app:min_PIRT}

The main limitation of the PIRT is that when rejecting the null hypothesis at significance level \( \alpha \), the probability of a false rejection can be as high as \( 2\alpha \) instead of \( \alpha \). While one way to address this is to reject the null hypothesis when the \textit{p}-value is below \( \alpha/2 \), a more conservative testing procedure inspired by \citet{wen2023residual} can be considered. The core idea behind this minimization-based PIRT is to compute a test statistic that reflects the worst-case scenario across all possible treatment assignments. Specifically, I define the test statistic as

\[
\tilde{T}(D^{obs}) = \min_{d \in \{0,1\}^N} T(Y(D^{obs}), D^{obs}, d),
\]

where the test statistic \( T \) is evaluated for each potential treatment assignment \( d \). Based on this, I define the \textit{p}-value as follows.

\begin{definition}[Minimization-based PIRT] \label{def:pvalue_min}
The minimization-based PIRT is an unconditional randomization test defined by \( \phi^{min}(D^{obs}) = 1\{pval^{min}(D^{obs}) \leq \alpha\} \), where \( pval^{min}(D^{obs}): \{0,1\}^N \rightarrow [0,1] \) is the \textit{p}-value function:
\[
pval^{min}(D^{obs}) = P(T(Y(D^{obs}), D, D^{obs}) \geq \tilde{T}(D^{obs})) \text{ for } D \sim P.
\]
Here, \( T(Y, d, d') \) represents the pairwise imputable statistic used to evaluate the hypothesis.
\end{definition}

To calculate this \textit{p}-value in practice, Algorithm \ref{algo:min} is applied. It computes the mean of $1+R$ draws, where $r = 0$ corresponds to $d = D^{obs}$.

\begin{algorithm}[ht]
  \SetKwInOut{Input}{Inputs}
  \SetKwInOut{Output}{Output}
  \SetKwInOut{Compute}{Compute}
  \Input{Test statistic $T = T(Y(d), d)$, observed assignment $D^{obs}$, observed outcome $Y^{obs}$, treatment assignment mechanism $P$, and size $\alpha$.} 
    \For{$r=1$ to $R$}{
     Randomly sample $d^r \sim P$, and store $T_r \equiv T(Y(D^{obs}), d^r, D^{obs})$.\\ Store $T^{obs}_r \equiv T(Y(D^{obs}), D^{obs},d^r)$.
    }
    \Compute{$\tilde{T}^{\star}(D^{obs}) = \min_{r=1,\dots, R}(T^{obs}_r)$}
  \Output{\textit{p}-value: $\hat{pval}^{min} = \frac{1+\sum^R_{r=1} 1\{T_r \geq \tilde{T}^{\star}(D^{obs})\}}{1+R}$.\\ Reject if $\hat{pval}^{min} \leq \alpha$.}
  \caption{Minimization-Based PIRT Procedure} \label{algo:min}
\end{algorithm}

In the toy example, shown in Table \ref{tab:PIRT}, $\tilde{T}(D^{obs}) = 1$; thus, $pval^{min} = 1/2$. The crucial distinction between minimization-based PIRT and PIRT is that minimization ensures size control, as demonstrated by Theorem \ref{theo:validity_min}.

\begin{theorem} \label{theo:validity_min}
Suppose the partially sharp null hypothesis \( H^{\epsilon_s}_0 \) holds. Then, the minimization-based PIRT, as defined in Definition \ref{def:pvalue_min}, satisfies \( \mathbb{E}_P[\phi^{min}(D^{obs})] \leq \alpha \) for any \( \alpha \in (0,1) \), where the expectation is taken with respect to \( D^{obs} \sim P \).
\end{theorem}

\paragraph{Proof of Theorem \ref{theo:validity_min}.}

To avoid confusion, denote $P_{D^{obs}}$ as probability respect to $D^{obs}$ and $P_{D}$ as probability respect to $D$.

Under the null $H^{\epsilon_s}_0$, by Proposition \ref{prop:validity} and setting $d=D$, $d^\prime=D^{obs}$, we have $T(Y(D), D^{obs}, D) = T(Y(D^{obs}), D, D^{obs})$. Hence, we have $\tilde{T}(D^{obs})=min_{d \in \{0,1\}^N}(T(Y(d), D^{obs}, d))$.

Then, by construction, $\tilde{T}(D^{obs}) \sim \tilde{T}(D) \leq T(Y(D^{obs}), D, D^{obs})$, and

\[pval^{min}(D^{obs})=P_D(T(Y(D^{obs}), D, D^{obs}) \geq \tilde{T}(D^{obs})) \geq P_D(\tilde{T}(D) \geq \tilde{T}(D^{obs})). \]

Therefore, 
\[P_{D^{obs}}(pval^{min}(D^{obs}) \leq \alpha) \leq  P_{D^{obs}}(P_D(\tilde{T}(D) \geq \tilde{T}(D^{obs})) \leq \alpha).\]

Let $U$ be a random variable with the same distribution as $\tilde{T}(D)$, induced by $P$. Denote its cumulative distribution function by $F_U$. We then have $P_D(\tilde{T}(D) \geq \tilde{T}(D^{obs}))= 1- F_U\{\tilde{T}(D^{obs})\}$, which is a random variable induced by  $D^{obs}\sim P(D^{obs})$. Hence, $P_D(\tilde{T}(D) \geq \tilde{T}(D^{obs})) = 1 - F_U(U)$, and by the probability integral transformation, $P_D(\tilde{T}(D) \geq \tilde{T}(D^{obs}))$ respect to $D^{obs}$ has a uniform $[0, 1]$ distribution under $H^{\epsilon_s}_0$. Thus, for any  $\alpha \in [0, 1]$,

\[P_{D^{obs}}(pval^{min}(D^{obs}) \leq \alpha) \leq  P_{D^{obs}}(P_D(\tilde{T}(D) \geq \tilde{T}(D^{obs})) \leq \alpha) \leq \alpha. \]

\qed

\paragraph{Handling a Large Number of Potential Treatment Assignments.} 
When $N$ is large, finding the minimum $\tilde{T}(D^{obs})$ across all possible treatment assignments can be computationally intensive. To ensure the validity of Algorithm \ref{algo:min} when dealing with a large number of units, optimization methods can be used to approximate $\tilde{T}^R(D^{obs})$ such that $\tilde{T}(D^{obs}) \geq \tilde{T}^R(D^{obs}) - \eta_R$ with probability $1-\eta$. The rejection level can then be adjusted to $\tilde{\alpha}$ such that $\alpha = \tilde{\alpha}(1-\eta) + \eta$, thereby ensuring validity. However, this approach introduces additional computational complexity.

An alternative strategy is to combine CRTs with PIRTs to reduce the space of potential treatment assignments. As discussed by \citet{athey2018} and \citet{zhang2023}, researchers often limit the assignment space to only those assignments with the same number of treated units as in the observed assignment. This two-stage approach--first defining the number of treated units and then performing testing within the reduced assignment space--remains valid. Using PIRTs in this context increases the set of focal units, potentially improving test power.

\section{Discussion on Some Extreme Cases} \label{app:empty}

\paragraph{Emptiness of Imputable Units Set.} 
The emptiness of the imputable units set depends on three factors: the distance being tested, the network structure, and the randomization design.

First, the target distance interacts with the network structure. If the distance \( \epsilon_s > \max_{i,j} G_{i,j} \), meaning it exceeds any existing distance in the network, then there will be no units in the imputable units set. In this case, additional data may be required to gain sufficient power for the test, or the target distance \( \epsilon_s \) could be reduced. For clarity in the following discussion, we focus on the case where \( \epsilon_s = 0 \).

Second, with \( \epsilon_s = 0 \), if all units in the sample are treated, the imputable units set will still be empty. To detect the existence of interference, a sufficient number of units beyond our target distance across various treatment assignments is necessary to achieve reasonable power.

\paragraph{Cases with an Undefined Comparison Group.} 

The distance being tested, network structure, and randomization design also influence whether one of the comparison groups is undefined. To highlight the core intuitions, we focus on the case where \( \epsilon_s = 0 \), implying that we are testing for the existence of interference and not all units are treated. Thus, some untreated units remain to conduct the test.

A general example is a network of couples, where exactly one unit in each pair is treated. In the example from the main text, with treatment assignments rotating across couples, the neighborhood units set may be empty. In practice, the test statistic must then assume a very high value for implementation.

More generally, for each assignment \( d \) and given \( \epsilon_c \), let \( \|\{i: d_i=1\}\| \) denote the number of treated units, \( \|\{i: d \in \mathcal{D}_i(0)/\mathcal{D}_i(\epsilon_c)\}\| \) the number of units in the neighborhood set, and \( \|\{i: d \in \mathcal{D}_i(\epsilon_c)\}\| \) the number in the control set. Whenever the number of non-imputable units is equal to or exceeds the number of neighborhood units, there may exist a pair of assignments \( (D^{obs}, D) \) such that the neighborhood units set is empty. This principle also applies to the control units set. Therefore, to ensure that both neighborhood and control sets are defined, we impose Assumption \ref{assum:network}.

\begin{assumption}[Regularization when \( \epsilon_s = 0 \)] \label{assum:network}
\[
\min \left\{ \min_{d \in \{0,1\}^N}\|\{i: d \in \mathcal{D}_i(0)/\mathcal{D}_i(\epsilon_c)\}\|, \min_{d \in \{0,1\}^N}\|\{i: d \in \mathcal{D}_i(\epsilon_c)\}\| \right\} > \max_{d \in \{0,1\}^N}\|\{i: d_i=1\}\|
\]
\end{assumption}

It is worth noting that \( \|\{i: d_i=1\}\| = N - \|\mathbb{I}(d)\| \) when \( \epsilon_s = 0 \). Therefore, Assumption \ref{assum:network} implies that the groups of interest occupy a large proportion of the population across all treatment assignments. This condition depends on \( \epsilon_c \), the network structure, and the experimental design. With Assumption \ref{assum:network}, we ensure all comparison groups remain non-empty across different potential assignments.

\begin{proposition}\label{prop:network}
Suppose Assumption \ref{assum:network} holds. For any \( D^{obs} \in \{0,1\}^N \), the pairwise imputable statistic \( T(Y(D^{obs}), D, D^{obs}) \neq \infty \) across different \( D \). 
\end{proposition}

\paragraph{Proof of Proposition \ref{prop:network}.}
We proceed by contradiction. Assume there exists a \( d \in \{0,1\}^N \) such that \( T(Y(D^{obs}), d, D^{obs}) = \infty \). Without loss of generality, suppose for any \( i \in \mathbb{I}(D^{obs}) \), \( d \notin \mathcal{D}_i(\epsilon_c) \). Then, for any unit \( j \in \{i: d \in \mathcal{D}_i(\epsilon_c)\} \), it must be the case that \( j \notin \mathbb{I}(D^{obs}) \), and hence \( j \in \{i: D^{obs}_i=1\} \).

Thus, we have
\[
\|\{i: d \in \mathcal{D}_i(\epsilon_c)\}\| \leq \|\{i: D^{obs}_i=1\}\|
\]

However, we have 
\[
\|\{i: d \in \mathcal{D}_i(\epsilon_c)\}\| \geq \min_{d \in \{0,1\}^N}\|\{i: d \in \mathcal{D}_i(\epsilon_c)\}\|
\]

and

\[
\|\{i: D^{obs}_i=1\}\| \leq \max_{d \in \{0,1\}^N}\|\{i: d_i=1\}\|.
\]

This implies that 

\[
\min_{d \in \{0,1\}^N}\|\{i: d \in \mathcal{D}_i(\epsilon_c)\}\| \leq \max_{d \in \{0,1\}^N}\|\{i: d_i=1\}\|,
\]

which contradicts Assumption \ref{assum:network}.

\qed

\section{Framework to Determine the Boundary of Interference} \label{app:multi}

Building on the PIRT framework, we can determine the boundary of interference by estimating a sequence of partially sharp null hypotheses at varying distances \( \epsilon_s \). This approach is useful for selecting a pure control distance or assessing the extent of interference based on distance. To this end, I consider a sequence of distance thresholds:
\[
\epsilon_0 < \epsilon_1 < \epsilon_2 < \dots < \epsilon_K < \infty,
\]
where \( K \geq 1 \) is chosen to include the settings introduced in previous sections. For instance, if the goal is to test for the existence of interference, one could set \( K = 1 \) with \( \epsilon_0 = \epsilon_s = 0 \) and \( \epsilon_1 = \epsilon_c \).

Using this sequence of distances, I can test a series of null hypotheses as defined in Definition \ref{def:gen}, where $\epsilon_s \in \{\epsilon_0, \dots, \epsilon_K\}$. However, it is important to note that not all distance levels will yield non-trivial power. First, there is a trade-off between the number of thresholds tested and the power of each test. While testing more thresholds provides a richer understanding of how interference varies with distance, it can reduce the power to detect interference, especially if certain threshold groups lack sufficient units. Based on simulation results, I recommend ensuring that each exposure level includes at least 20 units to maintain sufficient power at a significance level of $\alpha=0.05$.

Second, in some cases, $\epsilon_K$ may represent the maximum distance in the network, leaving no further room for $\epsilon_c$. Although it remains possible to test $H^{\epsilon_K}_0$, alternative approaches---such as adjusting for the number of nearby treated units, as suggested by \citet{hoshino2023}---may be needed to construct a test statistic with non-trivial power. For simplicity, this section will focus on testing $H^{\epsilon_k}_0$ for $k \leq K-1$.

Following Definition \ref{def:gen} of $H^{\epsilon_s}_0$, the multiple hypotheses under consideration exhibit a nested structure:

\begin{proposition}\label{prop:nested}
Suppose there exists an index \(\bar{K} \geq 0\) such that for any \(k \leq \bar{K} - 1\), the partially sharp null hypothesis \(H^{\epsilon_k}_0\) is false and \(H^{\epsilon_{\bar{K}}}_0\) is true. Then, \(H^{\epsilon_k}_0\) is true for any \(k \geq \bar{K}\).
\end{proposition}

\paragraph{Proof of Proposition \ref{prop:nested}.}
By Definition \ref{def:gen}, if \(H^{\epsilon_{\bar{K}}}_0\) is true, then \(Y_i(d) = Y_i(d^\prime)\) for all \(i \in \{1, \dots, N\}\) and any \(d, d^\prime \in \mathcal{D}_i(\epsilon^{\bar{K}})\).

Observe that for any \(i \in \{1, \dots, N\}\), by Definition \ref{def:distance_set},
\[
\mathcal{D}_i(\epsilon_0) \supset \mathcal{D}_i(\epsilon_1) \supset \dots \supset \mathcal{D}_i(\epsilon_K).
\]

Thus, for any \(k \geq \bar{K}\) and any \(d, d^\prime \in \mathcal{D}_i(\epsilon_{k}) \subseteq \mathcal{D}_i(\epsilon_{\bar{K}})\), it follows that \(Y_i(d) = Y_i(d^\prime)\) for all \(i \in \{1, \dots, N\}\). By Definition \ref{def:gen}, \(H^{\epsilon_k}_0\) is true for any \(k \geq \bar{K}\).  
\(\qed\)

Proposition \ref{prop:nested} implies that interference is bounded within a certain distance. Given this nested structure, I aim to develop an inference method that determines such boundaries by rejecting the null hypothesis up to a certain distance and failing to reject it beyond that point. However, in practice, situations may arise where $H^{\epsilon_k}_0$ cannot be rejected but $H^{\epsilon_{k+1}}_0$ is rejected. This could happen either because the test lacks power to reject the false null $H^{\epsilon_k}_0$ or due to multiple hypothesis testing errors, which lead to an erroneous rejection of the true null $H^{\epsilon_{k+1}}_0$. To mitigate the risk of over-rejecting true null hypotheses, I propose controlling the FWER.

\begin{definition}[FWER over all \(H^{\epsilon_k}_0\) for \(k = 0, \dots, K-1\)] \label{def:fwer}
Given a test \(\varphi: \{0,1\}^N \rightarrow \{0,1\}^K\), which maps the data to decisions for each hypothesis \(H^{\epsilon_k}_0\), the family-wise error rate (FWER) is defined as
\[
\text{FWER} = P\left(\exists k \geq \bar{K} \text{ such that } \varphi_k(D^{obs}) = 1, \text{ meaning that } H^{\epsilon_k}_0 \text{ is rejected}\right),
\]
where \(\bar{K} \geq 0\) is such that for any \(k \leq \bar{K}-1\), \(H^{\epsilon_k}_0\) is false, and for \(k \geq \bar{K}\), \(H^{\epsilon_k}_0\) is true.
\end{definition}

The definition of the FWER in Definition \ref{def:fwer} is motivated by the nested structure of \(H^{\epsilon_k}_0\), where the null hypothesis is true for any \(k \geq \bar{K}\). The critical issue is determining how to reject all the \(H^{\epsilon_k}_0\) hypotheses when identifying the boundary of interference, while still ensuring control over the FWER.

%Again, to ensure we are measuring interference in the distance of interest, I recommend test statistics that involve a comparison between units in the distance interval $(\epsilon_k, \epsilon_{k+1}]$ from treated units and the units in the distance interval $(\epsilon_K, \infty)$ from treated units. 

%Between the two choices of $\epsilon_c$, the power differences are minimal across all algorithms, suggesting that a misspecified $\epsilon_c$ mildly affects the performance of the testing procedure in this simulation exercise. However, since group $(0.2,\infty)$ is the only pure control group, comparing $(0,0.1]$ and $(0.2,\infty)$ may be preferable in practice. Overall, the power comparison for $H^0_0$ suggests using pairwise comparison-based PNRT with a rejection level of $\alpha$ or, more conservatively, $\alpha/2$ for testing, and employing the difference-in-mean between $(0,0.1]$ and $(0.2,\infty)$ as the test statistics.

\subsection{A Valid Procedure to Determine the Neighborhood of Interference} \label{sec:seq}

A major challenge in testing the extent of interference with respect to distance lies in addressing the issue of multiple hypothesis testing when conducting a series of tests to identify the neighborhood of interference. To manage the increased error rate arising from multiple tests, and drawing inspiration from \citet{MEINSHAUSEN2008} and Section 15.4.4 of \citet{lehmann2006}, I propose Algorithm \ref{algo:multi}.

\begin{algorithm}[ht]
  \SetKwInOut{Input}{Inputs}
  \SetKwInOut{Output}{Output}
  \SetKwInOut{Set}{Set}
  \Input{Test statistic $T = T(Y(d), d)$, observed assignment $D^{obs}$, observed outcome $Y^{obs}$, and treatment assignment mechanism $P$.} 
  \Set{$\hat{K}=0$.}
  \For{$k=0$ to $K-1$}{
    Test $H^{\epsilon_k}_0$ using the PIRT procedure and collect $pval^k$.\\  
    If $pval^k \leq \alpha$, set $\hat{K}=k+1$ and reject $H^{\epsilon_k}_0$. \\ 
    If $pval^k > \alpha$, break.
  }
  \Output{Significant spillover within distance $\epsilon_{\hat{K}}$.}
  \caption{Sequential Testing Procedure} \label{algo:multi}
\end{algorithm}

Algorithm \ref{algo:multi} is designed to control the FWER while leveraging the nested structure of sequential hypothesis testing. Unlike traditional multiple hypothesis testing procedures, such as the Bonferroni-Holm method, which require rejecting at a smaller level than $\alpha$, this algorithm maintains the significance level without adjustment, potentially increasing power compared to conventional methods \citep{MEINSHAUSEN2008}. Moreover, if the unadjusted \textit{p}-values increase as $k$ increases, indicating that interference diminishes with distance, there is no loss of power compared to not adjusting for multiple hypothesis testing, as we would naturally stop rejecting beyond a certain distance. When using the PIRT for each $k$, rejecting at the $\alpha/2$ level ensures size control. For the partially sharp null hypothesis $H^{\epsilon_k}_0$, a natural choice for $\epsilon_c$ is $\epsilon_{k+1}$. Theorem \ref{theo:validity_multi} guarantees the FWER control of Algorithm \ref{algo:multi}.

\begin{theorem} \label{theo:validity_multi}
The sequential testing procedure constructed by Algorithm \ref{algo:multi} controls the FWER at \(\alpha\).
\end{theorem}

\paragraph{Proof of Theorem \ref{theo:validity_multi}.}

Without loss of generality, consider the minimization-based PIRT below. The same proof holds when using the PIRT with a rejection level of $\alpha/2$. 

Suppose for any $k < \bar{K}$, $H^{\epsilon_k}_0$s are false and $H^{\epsilon_{\bar{K}}}_0$ is true. Then, by Algorithm \ref{algo:multi}, if there exist $k \geq \bar{K}$ such that $H^{\epsilon_k}_0$ is rejected, it must be the case that $H^{\epsilon_{\bar{K}}}_0$ is rejected. Thus, by Definition \ref{def:fwer},
\[FWER = P(pval^1 \leq \alpha, pval^2 \leq \alpha, \dots, pval^{\bar{K}} \leq \alpha) \leq P(pval^{\bar{K}} \leq \alpha) \leq \alpha\]
because $H^{\epsilon_{\bar{K}}}_0$ is true.
\qed

For example, suppose $K=2$ with $(\epsilon_0, \epsilon_1, \epsilon_2) = (0, 1, 2)$. Algorithm \ref{algo:multi} can be implemented in two steps. First, collect $pval^0$ for $H^0_0$ and reject $H^0_0$ if $pval^0 \leq \alpha$. If $H^0_0$ is not rejected, report that no significant interference was found. If $H^0_0$ is rejected, proceed to the second step, collect $pval^1$ for $H^1_0$, and reject $H^1_0$ if $pval^1 \leq \alpha$. If $H^1_0$ is rejected, report significant interference within distance 2; if $H^2_0$ is not rejected, report significant interference within distance 1.

\subsection{Rationale for Using the FWER}\label{sec:pure}

Controlling the family-wise error rate (FWER) is not the only option in multiple testing. As \citet{michael2008} suggests, a false discovery rate (FDR) control may be more suitable for exploratory analyses by allowing a small number of type I errors in return for greater power. An FDR-based approach could be explored in future work. However, when policymakers intend to apply a policy in distant regions under a positive interference effect, the more restrictive FWER control prevents overly optimistic conclusions about the interference boundary. In such settings, FWER provides a conservative distance threshold and better accounts for interference in expected welfare calculations.

This procedure also aids in identifying a pure control group by defining a ``safe distance'' \(\epsilon_c\), as mentioned in Section \ref{sec:test_stat}. One natural choice is \(\epsilon_K\), the greatest distance at which non-trivial testing power remains. However, researchers might reduce this distance to include more control units and boost power. Algorithm~\ref{algo:multi} offers a principled method for selecting \(\epsilon_c\) but the resulting value may be smaller than the true interference boundary due to the algorithm's conservative nature. Weighing these trade-offs is crucial when deciding whether to incorporate a pre-testing step.

\section{Incorporating Covariate Adjustment} \label{app:covar}

In practice, we often have access to covariates $X$, and incorporating this information is crucial for enhancing the power of tests, particularly when these covariates are predictive of potential outcomes (\citealp{ding2021}). Since the choice of test statistic does not affect the validity of the testing procedure for the partially sharp null hypothesis of interest, I propose three approaches for incorporating covariates in the analysis.

The first approach is PIRT with regressions. As illustrated in the main text, this method involves conducting the PIRT using regression coefficients from a simple OLS model as the test statistic. This OLS model includes a binary variable indicating whether a unit receives spillovers at a certain distance and known covariates, such as information about the neighborhood and social center points. A similar approach is discussed in \citet{Puelz2022}.

The second approach is PIRT with residual outcomes. The key idea here is to use the residuals from a model-based approach, such as regression with covariates of interest, rather than the raw outcome variables. I first obtain predicted values $\hat{Y}_i$ for the sample outcomes and then use the residuals, defined as the difference between observed outcomes and predicted values $\hat{e}_i=Y^{obs}_i-\hat{Y}_i$, for the PIRT procedures as the $Y$ defined in the main text. A similar approach for FRTs is proposed by \citet{rosenbaum2020design}, with detailed discussion in Sections 7 and 9.2 of \citet{bassefeller2018}.

The third approach is PIRT using pairwise residuals. In this method, for each pair of treatment assignments $(D^{obs}, D)$,  I conduct a regression with covariates within the imputable units set to transform the outcomes into residuals before testing and constructing the \textit{p}-values accordingly. This approach can be viewed as combining the first and second methods.

\subsection{Investigation on the Power of Incorporating Covariates}

In this investigation, we extend the potential outcomes described in Table \ref{tab:potential} by incorporating two covariates, \( X_1 \) and \( X_2 \). The new control potential outcomes, \( Y^C_i(\text{new}) \), are simulated based on the original control outcomes \( Y^C_i \) from Table \ref{tab:potential} as follows:
\[
Y^C_i(\text{new}) = 2 + 0.5 \times X_1 + 0.3 \times X_2 + Y^C_i,
\]
where \( X_1 \) is a binary covariate drawn from a Bernoulli distribution with parameter 0.5, and \( X_2 \) is a continuous covariate drawn from a standard normal distribution:
\[
X_1 \sim \text{Bernoulli}(0.5), \quad X_2 \sim \mathcal{N}(0, 1).
\]

It is important to note that only the control potential outcomes, \( Y^C_i \), are modified by these covariates. The remaining potential outcomes for treated units follow the same functional relationships as described in Table \ref{tab:potential}. By introducing \( X_1 \) and \( X_2 \), the control potential outcomes become more variable, reflecting the added noise from the covariates.

Next, I apply the three methods introduced earlier---PIRT with regressions, PIRT with residual outcomes, and PIRT using pairwise residuals---to construct the power curves. The simulation procedure remains consistent with that described in the main text, focusing on displacement effects and one-sided tests using non-absolute coefficients.

\begin{figure}[!htbp]
    \centering
    \caption{Power Comparison of Testing Methods for Different Covariate Adjustments}
    \label{fig:power_covar}
    \includegraphics[width=1\textwidth]{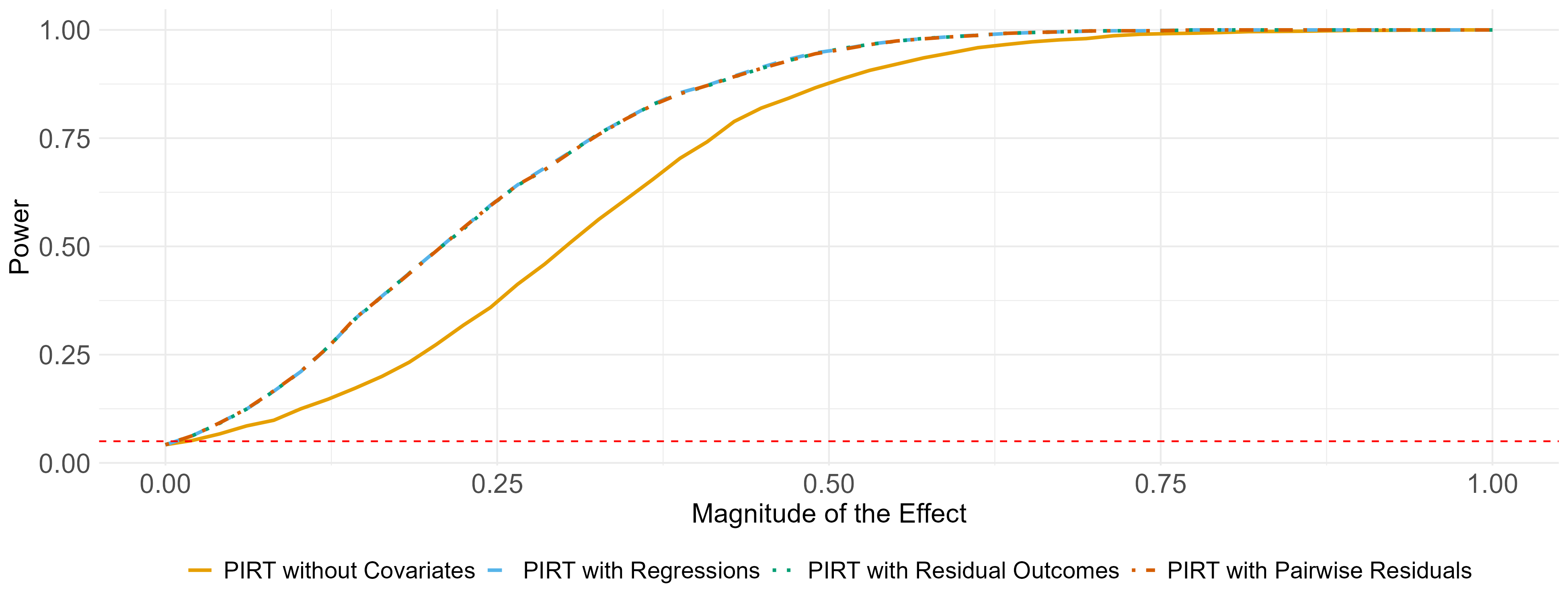}
    \notes{\textbf{Notes:} The red line indicates the size level $\alpha = 0.05$. Power is based on the PIRT with rejection at level $\alpha$. I consider 50 equally spaced values of $\tau$ between 0 and 1, conducting 2,000 simulations for each $\tau$ to compute the average rejection rate for each method.}
\end{figure}

Figure \ref{fig:power_covar} illustrates the power gains achieved by incorporating covariate information. While all methods involving covariate adjustments demonstrate similar power performance, leveraging covariates consistently results in higher power. When $\tau = 0$, the rejection rates for all methods align with the nominal size of the test. As $\tau$ increases, the power also increases. For instance, when $\tau \approx 0.25$, the power of the test with covariate adjustment reaches approximately 0.65, compared to less than 0.4 for the test without covariate adjustments. Therefore, in practice, researchers should select the method that best suits their specific context and data.

\subsection{Robustness of Results to Adjustment Methods}\label{app:covar_results}

The application of the above methods yields the results presented in Table \ref{tab:covar}. The regression models closely follow the framework outlined in \citet{blattman2021}, with slight modifications.

First, the regression includes the same covariates used in \citet{blattman2021}, such as police station fixed effects but excludes those related to the municipal services intervention.\footnote{The covariates include the following: number of crimes (2012–2015); average patrol time per day; square meters built (100 meters around) per meter of longitude; distance to the nearest shopping center, educational center, religious/cultural center, health center, and additional services office (e.g., justice); transport infrastructure (e.g., bus/BRT station); indicators for industry/commerce zones and service sector zones; income level; eligibility for municipal services; and interactions with the crime hotspot indicator.} In the original study, randomization testing was conducted jointly for both the policing and municipal services interventions, complicating interpretation when interaction effects are present. In this analysis, I hold the municipal services intervention fixed to isolate the effect of intensive policing.

Second, \citet{blattman2021} employs inverse propensity weighting in the weighted regression, using weights that account for both the hotspot policing and municipal services interventions.\footnote{Although this method does not fully eliminate bias, as discussed in \citet{aronow2020spillovereffectsexperimentaldata}, it helps address imbalance in the spillover group.} In this replication, I use weights that only consider the hotspot policing intervention to focus specifically on the impact of intensive policing.

\begin{table}[ht]
  \centering \caption{\textit{p}-Values: PIRT with Different Specifications}  \label{tab:covar}
  \setlength\tabcolsep{9pt}
    \renewcommand{\arraystretch}{1.2}
{\small \begin{tabular}{l|ccc}
  \hline \hline
   & \multicolumn{3}{c}{Unadjusted \textit{p}-values}     \\
      & $(0m, \infty)$ & $(125m, \infty)$ & $(250m, \infty)$   \\ 
  \hline
\textit{Violent crime} &  &  &     \\
Reg (WLS) & 0.105 & 0.719 & 0.158    \\ 
Reg (OLS) & 0.156 & 0.767 & 0.110    \\
Pair residuals & 0.119 & 0.726 & 0.142    \\
Residuals outcome & 0.114 & 0.757 & 0.166   \\
   \hline
\textit{Property crime} &  &  &     \\
Reg (WLS) & 0.508 & 0.232 & 0.619    \\ 
Reg (OLS) & 0.494 & 0.462 & 0.560    \\
Pair residuals & 0.481 & 0.252 & 0.565    \\ 
Residuals outcome & 0.455 & 0.250 & 0.578     \\
   \hline \hline
\end{tabular}  } 
\notes{ \textbf{Notes:} The table shows \textit{p}-values of PIRT across different methods, using the number of crimes as the outcome variable. Reg (WLS) is PIRT with regression, using the coefficient from the covariates-included regression with inverse propensity weighting as the test statistic. Reg (OLS) is the PIRT with regression, using the coefficient from the covariates-included regression without weighting as the test statistic. Pair residuals are PIRT with pairwise residuals, where residuals are constructed from the pairwise subset regression in the first step. The coefficient from the no-covariates regression with inverse propensity weighting is then used as the test statistic. Residuals outcome is PIRT with the residuals outcome, where residuals are constructed for all units in the first step, followed by using the coefficient from the no-covariates regression with inverse propensity weighting as the test statistic.} 
\end{table}

As shown in Table \ref{tab:covar}, the \textit{p}-values are very similar across the different methods, allowing researchers to choose the most practical implementation. Additionally, as discussed in Section C.3 of \citet{basse2024}, one can stratify potential assignments based on covariates to balance the focal units. This is done by stratifying both the permutations and the test statistic by an additional discrete covariate. However, we could not implement and compare \textit{p}-values from this method due to limitations in the original dataset.

In line with \citet{Puelz2022}, including covariates raises \textit{p}-values, suggesting that distance alone may not capture all heterogeneity in spillover effects.\footnote{Most of this \textit{p}-value increase stems from demographic covariates such as income levels and building density.} Covariates can reveal that the partially sharp null hypothesis does not fully account for unit-level heterogeneity. In an extreme scenario, if spillovers are perfectly correlated with these covariates, the partially sharp null would be rejected; however, regression adjustment could then eliminate the spillover signal, raising \textit{p}-values under the same null. Future research may refine distance measures by incorporating additional factors (e.g., socioeconomic disparities) to better capture spillover intensity \citep{Puelz2022}.

Researchers should interpret these results cautiously and decide on the null hypothesis of interest beforehand. If a researcher is interested in testing for no spillover effects after controlling for covariates, PIRTs can be extended to accommodate the work by \citet{Ding2016}. One can refer to \citet{owusu2023} for investigating heterogeneous effects in network settings. Alternatively, if interested in the weak null of the average effect being equal to zero (see \citet{ding2020}; \citet{basse2024}), one should note that the construction of \textit{p}-values in PIRTs differs from those in CRTs and FRTs, making classical approaches for weak nulls potentially inapplicable. Further investigation into these differences would be of interest to future research.

\section{Algorithm for Simulation Exercise in Section \ref{sec:simu}.} \label{app:algo}

\begin{figure}[!htbp] 
\centering
\caption{Unit Distribution}
\label{fig:dis_simu}
\includegraphics[width= 1 \textwidth]{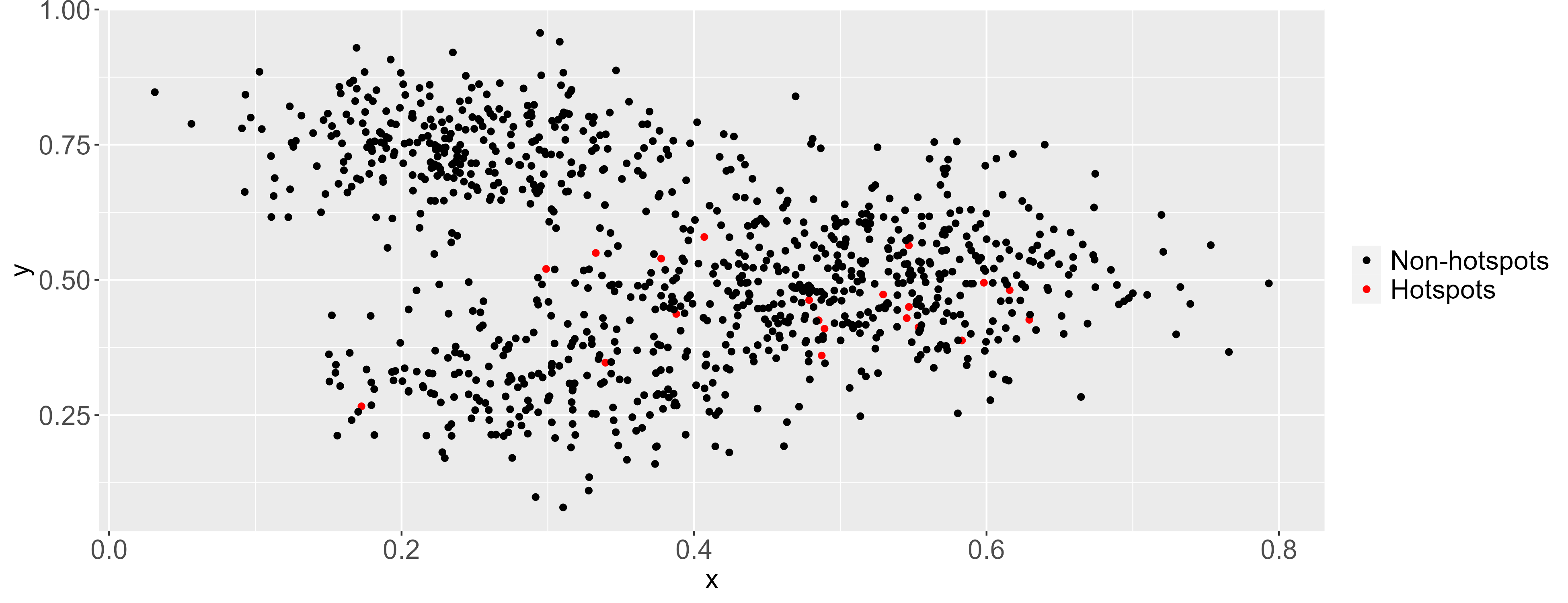}
\end{figure}

I generate $N = 1,000$ points from a bivariate Gaussian distribution with non-diagonal covariance to simulate the network on a $[0,1] \times [0,1]$ space. Figure \ref{fig:dis_simu} shows the unit distribution within this space. 

I focus on two distance thresholds, with $(\epsilon_0, \epsilon_1, \epsilon_2) = (0, 0.1, 0.2)$. Across different treatment assignments, the distance interval $(0, 0.1]$ comprises approximately 420 units, $(0.1, 0.2]$ around 250 units, and the pure control group $(0.2, \infty)$ around 320 units.

\begin{table}[ht] 
\centering 
\caption{Potential Outcome Schedule in the Simulation} 
\label{tab:potential} 
\renewcommand{\arraystretch}{0.935} {
\begin{tabular}{ll}
  \hline \hline
  Pure control for non-hotspots: & $Y^C_i \sim Gamma(0.086, 3.081)$\\
  Pure control for hotspots: & $Y^C_i \sim Gamma(0.737, 1.778)$\\
  Treated unit: & $Y^T_i = \max(Y^C_i - 1, 0)$\\
  Short-range spillover: & $Y_i(d) = Y^C_i +\tau \quad \forall d \in \mathcal{D}_i(0)/\mathcal{D}_i(0.1) $\\ 
  Long-range spillover: & $Y_i(d) = Y^C_i + 0.5\tau \quad \forall d \in \mathcal{D}_i(0.1)/\mathcal{D}_i(0.2)$\\
  \hline \hline
\end{tabular} }
\notes{ \textbf{Notes:} The outcome schedule is calibrated to the observed dataset. For $Gamma(k, \theta)$, $k$ is the shape parameter and $\theta$ is the scale parameter. $Y^C_i$ represents the pure control potential outcome for unit $i$, and $Y^T_i$ represents the potential outcome for unit $i$ when treated.} 
\end{table}

The algorithm for the simulation exercise in Section \ref{sec:simu} is outlined in Algorithm \ref{algo:simu}.

\begin{algorithm}[ht]
  \SetKwInOut{Input}{Inputs}
  \SetKwInOut{Output}{Output}
  \SetKwInOut{Set}{Set}
  \Input{5,000 randomly chosen assignments as the potential assignments set, $\mathbb{D}_S$.\\
  The biclique decomposition of $\mathbb{D}_S$ from \citet{Puelz2022}.}
  \Set{Spillover effect $\tau$ and corresponding schedule of potential outcomes.}
  \For{$s=1:S$}{
    Sample $D^{obs}_s$ from $\mathbb{D}_S$, and generate $Y^{obs}_s$.\\
    Implement the algorithms and collect corresponding $pval(D^{obs}_s)$ using $R=1,000$.
  }
  \Output{Average the number of rejections to obtain the power for that fixed $\tau$.}
  \caption{Simulation Study Procedure} \label{algo:simu}
\end{algorithm}

\end{appendices}
\end{document}